\documentclass[reprint,amsmath,amssymb,aps]{revtex4-2}

\usepackage{graphicx}
\usepackage{dcolumn}
\usepackage{bm}
\usepackage{hyperref}
\usepackage{lipsum}
\usepackage{subcaption}  
\usepackage{caption}         
\usepackage{float}
\usepackage{tikz}
\usepackage{ragged2e}
\usetikzlibrary{positioning, arrows.meta}  
\usepackage{relsize}  
\usepackage{overpic}  
\usepackage{placeins}
\usepackage{amssymb}
\usepackage{amsmath}
\usepackage{tabularx,array}
\usepackage[english]{babel}
\usepackage{xfrac}
\usepackage{soul}
\usepackage{xcolor}

\begin{document}

\preprint{APS/123-QED}

\title{Bridging the Analog and the Probabilistic Computing Divide:\\ Configuring Oscillator Ising Machines as P-bit Engines}

\author{E.M.Hasantha Ekanayake}
\affiliation{%
University of Virginia, Charlottesville, VA, USA
}%

\author{Nikhat Khan}
\affiliation{%
University of Virginia, Charlottesville, VA, USA
}%

\author{Nikhil Shukla}
\affiliation{%
University of Virginia, Charlottesville, VA, USA
}%

\begin{abstract}

Oscillator Ising Machines (OIMs) and probabilistic bit (p-bit) platforms have emerged as promising non-Von Neumann paradigms for tackling hard computational problems. While OIMs realize gradient-flow dynamics, p-bit platforms operate through stochastic sampling. Although traditionally viewed as distinct approaches, this work presents a theoretically grounded framework for configuring OIMs as p-bit engines. We demonstrate that this functionality can be enabled through a novel interplay between first- and second harmonic injection to the oscillators. Our work identifies new synergies between the two methods and broadens the scope of applications for OIMs beyond combinatorial optimization problems to those that entail stochastic sampling. We further show that the proposed approach can be applied to other analog dynamical systems, such as the Dynamical Ising Machine. 

\end{abstract}

\maketitle

\section{\label{sec:Intro} Introduction}

The endeavor to devise efficient solutions to complex computational problems has been a longstanding focus of science and technology research owing to its  far-reaching implications for practical applications. A particularly promising direction is the design of special-purpose hardware that accelerates such tasks while improving energy efficiency, especially for problems that remain challenging for conventional digital platforms. Within this context, two emerging approaches have attracted significant attention: analog oscillator Ising machines (OIMs) and probabilistic bit (p-bit) based computing engines-- the focus of the present work.

OIMs, first proposed by Wang \textit{et al.}~\cite{Wang2021} exploit the elegant equivalence between the 'energy function' characterizing the dynamics of a network of coupled oscillators under second harmonic injection (SHI) and the Ising Hamiltonian. Minimizing the Ising Hamiltonian is an archetypal combinatorial optimization problem (COP) where the goal is to find spin configurations $s \in \{-1,+1\}$ that minimizes the Ising Hamiltonian given by $H=-\sum{ J_{ij}s_i s_j}$, where $J_{ij}$ is the interaction between spin $i$ and $j$. OIMs realize gradient-flow dynamics on a continuous relaxation of the Ising model and have been extensively explored for solving combinatorial optimization problems (COPs)~\cite{mohseni2022ising,lucas2014ising}. They have been experimentally realized across diverse platforms, including optical~\cite{hamerly2019experimental,honjo2021100}, acoustic~\cite{litvinenko202550}, electronic~\cite{mallick2021overcoming,moy20221,bashar2020experimental,vaidya2022creating,maher2024cmos,cilasun2025coupled}, spin-wave~\cite{litvinenko2023spinwave}, and quantum systems~\cite{king2022coherent}. These demonstrations have been complemented by extensive theoretical studies~\cite{Bashar20232,cheng2024control,allibhoy2025global}.

P-bit-based computing engines, on the other hand, are a complementary computing paradigm uniquely capable of Boltzmann sampling. A p-bit can be viewed as a tunable random number generator whose output probability depends on the synaptic input. At the network level, this yields a binary stochastic neural network (BSNN)~\cite{camsari2019p}, with spin updates governed by,
\begin{equation}
    s_i^{+} = \text{sgn}\left[ \tanh \left( \beta \sum_{\substack{j=1 \\ j \ne i}}^{N} J_{ij} s_j  \right) - \mu\right] \label{pbit1}
\end{equation}

where, $\mu$ is a random number typically selected from a uniform distribution between $[-1,1]$, and $\beta$ is the equivalent of inverse temperature  \cite{camsari2017stochastic}. Furthermore, if we interpret the spins as the phases of oscillators—a perspective that is useful in the context of this work—then the state update rule can be expressed as,
\begin{equation}
    s_i^{+} = \cos\phi_i= \text{sgn}\left[ \tanh \left( \beta \sum_{\substack{j=1 \\ j \ne i}}^{N} J_{ij} \cos\phi_j  \right) - \mu\right] \label{pbit2}
\end{equation}
\noindent where, $\phi \in \{0,\pi\}$ (wrapped phase form). In both Eqs. \eqref{pbit1} and \eqref{pbit2}, the self bias term has not been considered although it is relatively straightforward to include it into the approach presented here.

The intrinsic stochastic sampling capability of p-bit engines offers a complementary approach to OIMs in the context of solving COPs~\cite{aadit2022massively,chowdhury2023full,whitehead2023cmos,duffee2024integrated,borders2019integer,singh2024cmos,si2024energy,jhonsa2025cmos}. In addition, the stochastic sampling can be exploited in other application such as probabilistic inference and learning. Establishing such a capability within OIMs—hitherto not demonstrated—could therefore significantly broaden their functional scope beyond combinatorial optimization. To date, OIMs and p-bit engines have largely been pursued as independent approaches, with limited exploration of their potential synergies~\cite{bohm2022noise,lee2025noise}. In this work, we bridge this gap by demonstrating the feasibility of configuring OIMs as p-bit engines, thereby establishing them as an alternative analog platform for probabilistic computing.

\begin{figure*}[htbp]
    \centering
    \includegraphics[width=0.8\linewidth]{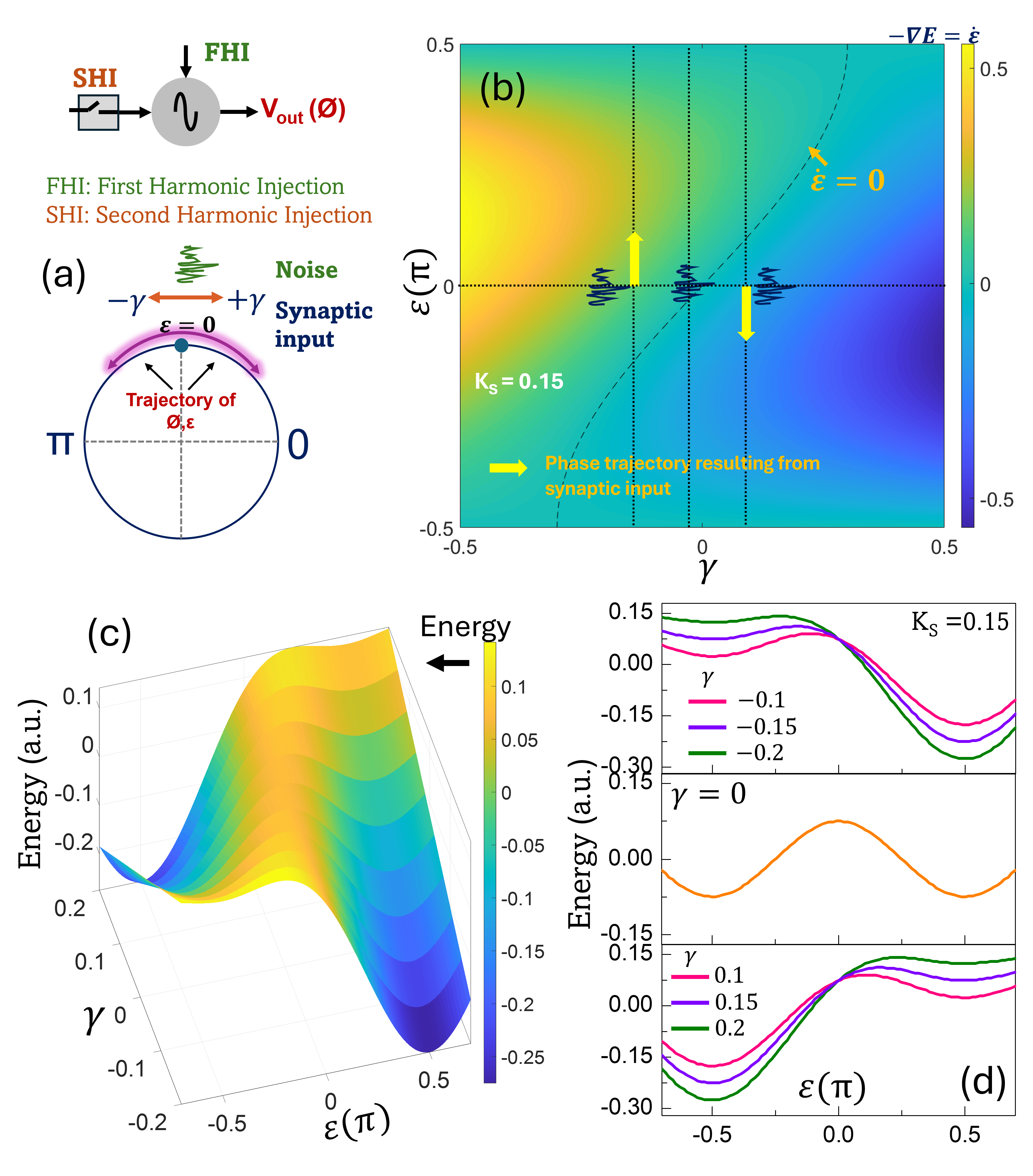}
    \caption{\textbf{Dynamics of a harmonic oscillator under SHI} (a) Schematic illustration of the signals required to program an harmonic oscillator as a p-bit. (b) Force field as a function of $\gamma$ ($K_s=0.15$, where $K_s$ is the strength of SHI.). (c) Corresponding energy landscape demonstrating its evolution with the synaptic input, $\gamma$. (d) Specific cuts of the energy landscape at $\gamma=\{0,\,\pm\,0.1,\,\pm\, 0.15,\,\pm\,0.2\}$ ($K_s=0.15$).}
    \label{fig:Fig.1}
\end{figure*}
To establish the foundation of our oscillator-based p-bit engine, we begin by demonstrating two key properties of harmonic oscillators (the class of harmonic oscillators considered in this study) and their networks operating under SHI: (a) An oscillator subjected to first- and second harmonic injection can function as a binary stochastic neuron (BSN); and (b) A network of such coupled oscillators—specifically an OIM—can operate as a binary stochastic neural network (BSNN). These two properties form the conceptual and mathematical basis for designing an  oscillator-based p-bit engine.

\section{\label{sec:BSN}Configuring Oscillators as Stochastic Neurons}

To design a binary stochastic neuron (BSN) using an oscillator, we first examine the dynamics of a harmonic oscillator subjected to two external inputs (Fig.~\ref{fig:Fig.1}(a)). \textbf{(a)} The first input is a signal with a frequency nearly equal to the oscillator’s natural frequency—a condition commonly referred to as injection locking. Within a specific locking range, this external signal entrains the oscillator, steering its output toward the phase and frequency of the injected signal, effectively synchronizing the oscillator's behavior with the input. An energetics-based explanation of the relevant phase behavior is provided in Appendix~\ref{appendix1}. We refer to this signal as the fundamental harmonic injection (FHI).  All phase values are defined with respect to a common reference signal. \textbf{(b)} The second input is SHI, operating at twice the oscillator’s natural frequency. SHI drives the oscillator phase toward either $\phi = 0$ or $\phi = \pi$.

The resulting phase dynamics of such an oscillator can be described by,

\begin{equation}
\begin{split}
\frac{d\phi}{dt}&=-K_c\sin{\left(\phi-\theta\right)}-K_s\sin(2\phi) \\
\label{neuron1}
\end{split}
\end{equation}

\noindent where, $\phi$ denotes the output phase of the oscillator, while $\theta$ represents the phase offset of the FHI signal input. The parameters $K_c$ and $K_s$ are coupling constants of the FHI and SHI signals, with the first and the second terms on the RHS of Eq. \eqref{neuron1} capturing the influence of the FHI and SHI, respectively.

In this work, we will restrict our attention to cases where $\theta \in \{0, \frac{\pi}{2}, \pi\}$. Initially, we will focus on the analysis where $\theta \in \{0, \pi\}$, with the case $\theta = \frac{\pi}{2}$ becoming relevant further on. Under this constraint, we can recast Eq. \eqref{neuron1} as,
\begin{equation}
\begin{split}
\frac{d\phi}{dt}=-\gamma\sin{\left(\phi\right)}-K_s\sin(2\phi) 
\label{neuron2}
\end{split}
\end{equation}
\noindent where $\gamma \,(=\pm K_c)$ denotes the scaled synaptic input; $\gamma = K_c$ when $\theta = 0$ and $\gamma = -K_c$ when $\theta = \pi$. For simplicity, we will generally refer to $\gamma$ as the synaptic input in the following discussion. The details of the scaling factor will be elaborated in the subsequent section. 

Furthermore, for convenience, we perform a frame rotation by \(0.5\pi\), redefining the phase as $\phi=\frac{\pi}{2}+\epsilon$. We also note that the relevant values of $\theta$ for this rotated frame shift to $\theta^{'} \in \{-\frac{\pi}{2}, 0, \frac{\pi}{2}\}$. Under this transformation, equation \eqref{neuron2} becomes,

\begin{equation}
\frac{d\epsilon}{dt}=-\gamma\cos{\left(\epsilon\right)}+K_s\sin(2\epsilon)  \label{neuron3}
\end{equation}

We now analyze the properties and the behavior of Eq.~\eqref{neuron3}. First, the fixed points of this system lie at $\epsilon_1^*=\pm0.5\pi$, and at values satisfying $\sin(\epsilon_2^*)=\frac{\gamma}{2K_s}$. A detailed stability analysis of all the fixed points has been presented in Appendix \ref{appendix2}.

Next, we investigate the expected dynamical behavior of the system by analyzing the effective force field \((-\nabla E=\dot{\epsilon})\) driving the phase evolution as a function of \( \gamma \), as shown in Fig.~\ref{fig:Fig.1}(b). 
The dotted line in the figure represents the set of phase points where the phase velocity vanishes, i.e., \( \frac{d\epsilon}{dt} = 0 \), and is described by the equation:
\[
-\gamma + 2K_s \sin(\epsilon) = 0
\]
This curve defines the \emph{nullcline} of the system, separating regions of positive and negative phase flow.

Interestingly, when the system is initialized at \( \epsilon = 0 \), the direction of phase evolution depends entirely on the sign of \( \gamma \). For \( \gamma < 0 \), the dynamics flow toward \( \epsilon = \frac{\pi}{2} \) (corresponding to \( \phi = \pi \)). Conversely, for \( \gamma > 0 \), the dynamics flow toward \( \epsilon = -\frac{\pi}{2} \) (i.e., \( \phi = 0 \)). Moreover, the magnitude of \( \gamma \) determines the steepness of the phase flow at $\epsilon=0$. At the critical point \( \gamma = 0 \), the direction of phase evolution becomes entirely stochastic in the presence of noise.

Figure~\ref{fig:Fig.1}(c) shows the evolution of the oscillator's energy landscape with the synaptic input, \( \gamma \), with Fig.~\ref{fig:Fig.1}(d) showing cuts at specific values of $\gamma=\{0,\,\pm\,0.1,\,\pm\, 0.15,\,\pm\,0.2\}$. In Fig.~\ref{fig:Fig.1}(d), the relative symmetry about \( \epsilon = 0 \) is evident across all cases, with the energy profiles for \( \gamma = \{+0.1,\,+ 0.15,\,+ 0.2\} \) and \( \gamma = \{-0.1,\,-0.15,\,- 0.2\} \) appearing as mirror images, respectively, and the \( \gamma = 0 \) case exhibiting a perfectly symmetric landscape. 

In alignment with prior work on magnetic tunnel junctions (MTJs)-based p-bits ~\cite{faria2017low}, we engineer our oscillator-based BSN to operate within the low energy barrier regime. Since the height of the barrier is controlled by $K_s$ (see Appendix \ref{appendix3} for details), this is achieved by tuning $K_s$ to a small value ($K_s\ll1$) during the sampling event. Thus, the oscillator-based approach can enable a BSN with a tunable barrier.

From \ref{fig:Fig.1}(b) as well as from the form of Eq. \eqref{neuron3}, it can be observed that the system exhibits a unique symmetry at \( \epsilon = 0 \), within the domain \( \epsilon \in \left[-\frac{\pi}{2}, \frac{\pi}{2}\right] \), whereby the magnitude of the (scaled) synaptic input has a symmetric effect for \( +\gamma \) and \( -\gamma \), but induces phase flows in opposite directions. This symmetry plays a critical role in enabling the oscillator to function as a BSN, as it defines a neutral point from which the system can stochastically evolve toward one of two stable states depending on the sign of the synaptic input. As we will show later, this stochastic response is also highly non-linear. In practical settings, the oscillator can be driven to this neutral point \( \epsilon = 0 \) using an FHI signal with a phase offset of \( \theta^{'} = 0 \) (we refer to this input as \(\text{FHI}^0\)), and without applying the SHI.

Thus, operating the oscillator as a BSN entails the following steps:\\

\noindent \textbf{(i)} Set oscillator phase to the neutral point using \(\text{FHI}^0\).\\\\
\textbf{(ii)} Remove \(\text{FHI}^0\) followed by application of the  synaptic input (also an FHI signal when considering a single oscillator) and SHI signal (applied as a ramp) to evaluate the (stochastic) neuron's state---this constitutes a sampling event. As we will demonstrate later, in the OIM network, such synaptic input is effectively generated by other connected oscillators within the network.

\section{\label{sec:oBSN}Dynamics of an Oscillator-based BSN}\
To quantitatively analyze the stochastic nonlinear response of the oscillator-based BSN, we begin with the oscillator dynamics described in Eq.~\eqref{neuron3}. We first define the updated spin state in terms of $\epsilon$, which, in the context of the continuous time dynamics considered here, is given by,
\[
s^+=\text{sgn}\left(\cos(\phi^+)\right)=-\text{sgn}\left(\sin(\epsilon^+)\right)
\]
where, \(\phi^+\) and \(\epsilon^+\) refer to the phase of the system at a small time increment \(\Delta t \rightarrow0\) after the sampling has been initiated.

We now evaluate the solution to the dynamics presented in Eq.~\eqref{neuron3}. Although deriving an explicit solution is challenging, Eq.~\eqref{neuron3} has an implicit analytical solution given by,
\begin{equation}
    \frac{\zeta(\epsilon) - \gamma \tanh^{-1}(\sin(\epsilon))}{\gamma^2-4K_s^2}=t \label{BSN_1}
\end{equation}
\noindent where,

\begin{equation}
\begin{split}
    \zeta(\epsilon) &= K_s \big( -2 \log(\gamma - 2K_s \sin(\epsilon)) \\
    &+ \log(1 - \sin(\epsilon)) + \log(\sin(\epsilon) + 1) \big)\\                    \notag   
    \end{split}
\end{equation}
For it's applications as a BSN, we focus on the direction of the initial phase trajectory. To understand and evaluate the system dynamics at the sampling instant—defined as the moment when (FHI\(^0\)) is suppressed and the SHI and the synaptic input are asserted —we adopt the following approximation:\\\\
\noindent \textbf{(i)} The dynamics are evaluated in the limit $t\rightarrow0$.\\\\
\textbf{(ii)} We model the noise as Gaussian white noise with
\(\langle \eta(t) \rangle = 0, \quad \langle \eta(t)\eta(t') \rangle = 2K_n \delta(t - t'),\) with $K_n$ denoting the noise intensity
i.e., \(\eta(t)\,dt = \sqrt{2K_n}\,dW_t,\) where \(W_t\) is a Wiener process. Equivalently, over a finite timestep $\Delta t$,
\(\int_{t}^{t+\Delta t} \eta(\tau)\,d\tau \;\sim\; \mathcal{N}\!\big(0,\,2K_n\,\Delta t\big).\)\\\\
\textbf{(iii)} At the onset of sampling ($t\rightarrow0$) , we assume \( K_s \rightarrow0\) such that \(K_s\ll |\gamma| \). This reflects the requirement for a low energy barrier in the probabilistic regime. Subsequently, $K_s$ must be ramped up for reasons discussed in the following section.

Under these approximations, the oscillator phase can be expressed as:

\begin{equation}
\begin{split}
\label{BSN_2}
\sin(\epsilon_i)&\approx\tanh\left(-\gamma t+\epsilon^\eta(t) \right)\\\\\
&\approx\tanh\left(-\gamma  t\right)+\epsilon^{\eta}(t). \text{sech}^2(\gamma  t)\\\\
\end{split}
\end{equation}

\noindent Here, $\epsilon^\eta(t)\sim \mathcal{N}\!\big(0,\,2K_n\,t\big)$, and is small such that the $\tanh(.)$ term can be linearized. Accordingly, at a small time increment \(\Delta t\rightarrow0\) after the onset of sampling at t=0, the updated state of the oscillator-based BSN can be described as,
\begin{equation}
\begin{split}
s^+ &\approx \text{sgn}\left[ \tanh(\gamma \Delta t) - \epsilon^\eta(\Delta t) .\text{sech}^2(\gamma \Delta t) \right]\\\\
&\equiv \text{sgn}\left[ \tanh(\gamma \Delta t) - \vartheta \right]
\label{BSN_3}
\end{split}
\end{equation}
\noindent where, $\vartheta\equiv \epsilon^\eta(\Delta t) .\text{sech}^2(\gamma \Delta t)$. Since $\text{sech}^2(\gamma \Delta t) \in (0,1]$, and noise intensity is assumed to be small, $\vartheta$ has a high probability of being in the interval [-1,+1]. The infinitesimal time ($\Delta t$) can be interpreted as the time duration required by the oscillator phase to reach a critical threshold magnitude $|\epsilon_{th}|$, beyond which the oscillator phase cannot ‘reverse’ trajectory.  $\Delta t$ can be tuned through properties of the SHI pulse, such as its slew rate. Additional details on $\Delta t$ are discussed in the following sections. The derivation of Eqs.~\eqref{BSN_2} and~\eqref{BSN_3} has been presented in Appendix ~\ref{appendix4}. Furthermore, while Eq.~\eqref{BSN_2} addresses the regime where $\gamma \gg K_s$, we also consider (in Appendix ~\ref{appendix4}) the complementary case where both \(\gamma\) and \(K_s\) are small and comparable. In this setting, the synaptic bias $\gamma$ and the stochastic perturbations act as competing drivers of the phase dynamics. A large $|\gamma|$ produces a predictable exponential drift, whereas strong noise leads to rapid amplification of fluctuations and broad dispersion of trajectories. The observed evolution is therefore governed by the balance between deterministic drive and stochastic forcing.

We note that the actual synaptic input to the BSN would be applied as the voltage (or current) amplitude, \(V_{\text{inj}}\), and phase (in-phase or out-of-phase) of the injected signal, which relate to \(\gamma \) and the coupling constant \(K_c\) as \(\gamma = \pm\, K_c\, \approx \pm\,\, \frac{\omega_0}{2Q} \cdot \frac{V_{\text{inj}}}{V_{\text{osc}}}\) \cite{adler2006study,bhansali2009gen}, where $Q$ is the quality factor of the oscillator, $\omega_0$ is the natural frequency, and \(V_{\text{osc}}\) represents the amplitude of the oscillator, respectively. This formulation implies that \(\gamma\) serves as a scaled representation of the synaptic input, as noted earlier. Furthermore, the effective inverse temperature is then given by \(\beta_{\text{eff}} = \frac{\omega_0 \Delta t}{2 V_{\text{osc}}}.\frac{1}{Q}\) implying that \(\beta_{\text{eff}}\) can be modulated using the quality factor, $Q$, of the oscillator as well as other parameters such as the oscillation amplitude among others. This tunability provides a practical mechanism for controlling the stochastic behavior of the system. The updated state is then expressed as

\begin{equation}
\begin{split}
s^+ &\approx \text{sgn}\left[ \tanh\left( \frac{\omega_0 \Delta t}{2Q V_{\text{osc}}} \cdot V_{\text{inj}} \right) - \vartheta \right]  \\\\  \label{final_BSN}
&\equiv   \text{sgn}\left[ \tanh\left( \beta_{\text{eff}} \cdot V_{\text{inj}} \right) - \vartheta \right]
\end{split}
\end{equation}

\begin{figure}[htbp]
    \centering
    \includegraphics[width=1\linewidth]{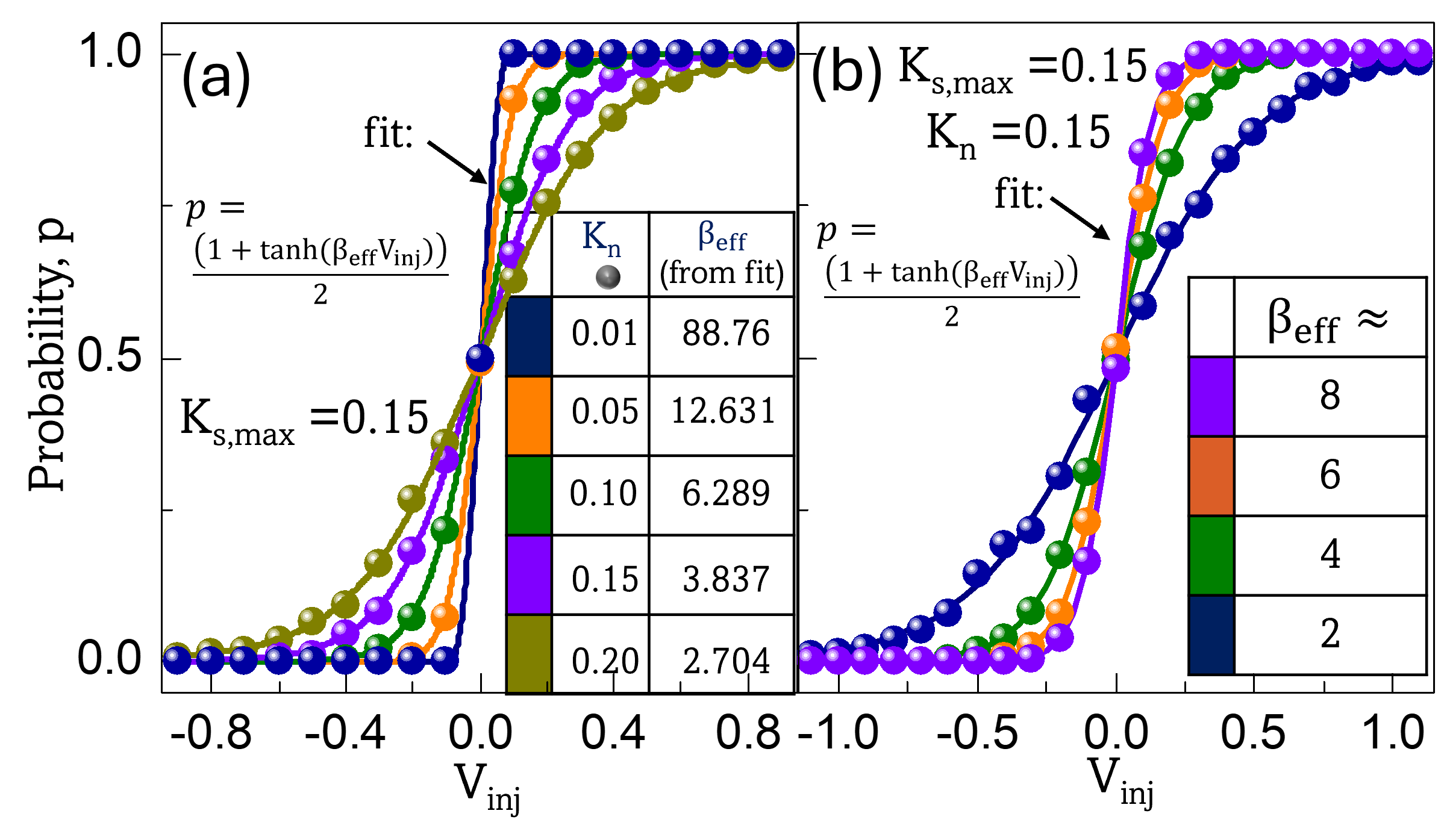}
    \caption{\textbf{Oscillator-based BSN.} Firing probability (symbols) as a function of the synaptic input for: (a) varying levels of noise ($K_n$). (b) different values of $\beta_{\text{eff}}$ derived for $K_n=0.15$. We note that the $\beta_{\text{eff}}$ profile will change for a different $K_n$. The lines in the plot indicate fits using the equation $p=\frac{1+\tanh(\beta_{\text{eff}}V_{\text{inj}})}{2}$ along with the calculated $\beta_{\text{eff}}$; $p:$ firing probability. All fits exhibit $R^2>0.999$}
    \label{fig:Fig.2}
\end{figure}

Equation~\eqref{final_BSN} showcases the BSN's capability to perform Boltzmann sampling, and consequently, function as a p-bit when initialized at the critical phase point, \( \epsilon = 0 \). In the following section, we show that, in OIMs, \(V_{\text{inj}}\)—and consequently \(\gamma\)—is generated and regulated through the mutual feedback among the coupled oscillators. To validate the dynamics derived above, we simulate the oscillator-based BSN's switching using a stochastic differential equation solver implemented in MATLAB\textsuperscript{\textregistered}. We first examine the evolution of stochastic behavior under varying noise levels. The phase is initialized at \(\epsilon = 0\) and allowed to evolve in the presence of different noise intensities and synaptic inputs. The firing probability is then estimated over 2000 such cycles. Figure~\ref{fig:Fig.2}(a) presents the simulation results (symbols) for the output firing probability of the oscillator-based BSN as a function of \(V_{\text{inj}}\). These results are fitted using the function \(p = \frac{1 + \tanh(\beta_{\text{eff}} V_{\text{inj}})}{2} \) (solid lines), showing excellent agreement with the simulated data (\(R^2 > 0.999\)); p is the probability of the neuron firing i.e., switching to $s=+1$. 

In practical implementations, modulating external noise to control temperature and stochasticity may not be feasible. Instead, the effective inverse temperature—and thus the degree of stochastic behavior—can be tuned by adjusting \(\beta_{\text{eff}}\), which is achievable through modulation of the oscillator’s quality factor \(Q\). Figure~\ref{fig:Fig.2}(b) illustrates the evolution of firing probability (symbols) for different values of \(\beta_{\text{eff}}\), where the switching probability again exhibits tanh(.) dependence on \(V_{\text{inj}}\) as confirmed by the corresponding fitted curves (solid lines) with \(R^2 > 0.999\). A noise strength of \(K_n=0.15\) was used in the simulation. These simulation results further support that the oscillator exhibits the characteristic behavior of a BSN capable of performing Boltzmann sampling.

A key consideration in engineering the oscillator's dynamics for BSN functionality is the relative magnitude of \( \gamma \) and the SHI strength, \( K_s \). Realizing effective Boltzmann sampling behavior requires a small energy barrier, which corresponds to the regime \( K_s \rightarrow 0 \) such that \( K_s \ll |\gamma| \). If this condition is not satisfied, the system may still operate as a BSN; however, its dynamics may deviate from the Boltzmann sampling behavior. This is detailed in the analysis in Appendix \ref{appendix4}.

In contrast, to preserve the oscillator's phase trajectory after sampling and to enable reliable readout, a lower bound on \( K_s \) must be satisfied. Specifically, this bound ensures that once the phase magnitude exceeds a certain threshold (\(|\epsilon_{\text{th}}|\)), the phase continues to evolve in the same direction until it reaches the corresponding fixed point. 

While a detailed analytical derivation is provided in Appendix~\ref{appendix5}, we offer here a qualitative explanation of the origin of this constraint by examining the dynamics described by Eq.~\eqref{neuron3}. Within the interval \( \epsilon \in (-\frac{\pi}{2}, \frac{\pi}{2}) \), the cosine term satisfies \( \cos(\epsilon) > 0 \), and the synaptic input term \( -\gamma \cos(\epsilon) \) therefore drives the phase evolution in the direction \emph{opposite} to the sign of \( \gamma \). This implies that the synaptic input alone tends to push the phase toward the fixed point opposite to the sign of \( \gamma \), unless counteracted by the SHI term.

If noise initially drives \( \frac{d\epsilon}{dt} \) in the direction not favored by the synaptic input (Fig.~\ref{fig:Fig.6}(b)), and \( K_s \) is too small, the oscillator may reverse its trajectory \emph{to align with the trajectory favored by the synaptic input}. In such cases, the final state may not reflect the oscillator’s initial trajectory or the intended output \( s^+ \). The SHI term--- specifically, the \( K_s \sin(2\epsilon) \) component in Eq.~\eqref{neuron3}, acts to reinforce the initial direction of \( \frac{d\epsilon}{dt} \) \emph{generated by the stochastic sampling process}, provided \( K_s \) is sufficiently large. This reinforcement helps ensure that the phase continues toward the correct fixed point.

The critical condition to ensure that an oscillator at \( \epsilon = |\epsilon_{\text{th}}| \) maintains its trajectory is given by:
\[
|\gamma| < 2K_s \sin(|\epsilon_{\text{th}}|)
\]
It is important to note that due to the presence of noise, this threshold is inherently probabilistic, resulting in a diffuse rather than deterministic boundary. This condition appears to contradict the earlier requirement concerning the relative magnitudes of \( \gamma \) and \( K_s \). To reconcile these seemingly opposing constraints, we propose implementing the SHI input as a ramp signal with a carefully engineered slew rate. Specifically, the SHI can be initialized at a low amplitude—ensuring that \( K_s \ll |\gamma| \)—to facilitate stochastic sampling in the low-barrier regime. Subsequently, the amplitude of the SHI is increased to reinforce the resulting phase trajectory. Additionally, as noted above the characteristics of the applied SHI input, particularly its slew rate, also modulate $\Delta t$. A smaller slew rate increases $\Delta t$, which in turn modifies the inverse temperature and also makes the system less stochastic (see Appendix~\ref{appendix:slew_rate} for details).

The analysis presented in this section reveals that the competition between synaptic input and noise gives rise to a stochastic bifurcation, which constitutes the core operating principle of the proposed oscillator-based BSN.  

\section{Configuring OIM\MakeLowercase{s} As Binary Stochastic Neural Networks} \label{sec:OIM}

Building upon the ability to configure an oscillator as a BSN capable of performing Boltzmann sampling, we now investigate the possibility of configuring an OIM as a p-bit engine, or in other words, a BSNN. The core premise of this idea is that \emph{after} the system reaches steady state, i.e., \( \epsilon \in \{-\frac{\pi}{2}, \frac{\pi}{2}\} \), the dynamics of a randomly sampled oscillator driven to the phase point \( \epsilon = 0 \) (i.e., \( \phi = \frac{\pi}{2} \)) still approximate Boltzmann sampling. Steady state initialization at \( \epsilon \in \{-\frac{\pi}{2}, \frac{\pi}{2}\} \) can be achieved by applying an SHI signal prior to the onset of stochastic sampling. As noted earlier, in the OIM, the feedback from other coupled oscillators acts as the effective synaptic input, which subsequently modulates the oscillator's stochastic dynamics.

To establish this result, in the following sections, we divide the oscillators in the network into two categories and analyze their dynamics:\\\\

\noindent \textbf{(i) Randomly sampled oscillator $i$ initialized to \( \mathbf{\epsilon_i = 0} \)}.\\\\
The phase evolution of an oscillator in the OIM network can be described by the equation,
\[
\frac{d\phi_i}{dt} = -K \sum_{\substack{j=1 \\ j \ne i}}^N J_{ij} \sin(\phi_i - \phi_j) - K_s \sin(2\phi_i)\\ 
\]
which, in the rotated frame of reference, can be expressed as,\\
\begin{equation}
\frac{d\epsilon_i}{dt} = -K \sum_{\substack{j=1 \\ j \ne i}}^N J_{ij} \sin(\epsilon_i - \epsilon_j) + K_s \sin(2\epsilon_i)
\label{s_OIM1a}
\end{equation}

\noindent As alluded to earlier, we assume that:\\

\noindent(i) The selected oscillator is initialized at $\epsilon=0$. This can be accomplished using \(\text{FHI}^0\) signal with large amplitude.\\\\
(ii) Since we begin performing stochastic sampling after the OIM network has achieved steady state, all \emph{other} oscillators are at $\epsilon=\pm \frac{\pi}{2} \:\: \big(\phi \in \{0,\pi\}\big) $. Furthermore, we will ensure that the oscillators maintain their state during the sampling event by applying a sufficiently large $K_s$. This condition is required to ensure that the oscillator dynamics map to Gibbs sampling (see Appendix~\ref{appendix:non_sampled_oscillators}). In the subsequent sections, we will discuss how these conditions can be implemented. 

Under the constraints outlined above, the phase dynamics of the selected oscillator simplify to:
\begin{equation}
\begin{split}
\frac{d\epsilon_i}{dt} &=
 \left(K\sum_{\substack{j=1 \\ j \ne i}}^N J_{ij} \sin(\epsilon_j)\right)\cos(\epsilon_i) + K_s \sin(2\epsilon_i) \\\\
&\equiv -\gamma_i \cos(\epsilon_i)+ K_s \sin(2\epsilon_i) 
\label{s_OIM2}
\end{split}
\end{equation}

where,
\[
\gamma_i=-K\sum_{\substack{j=1 \\ j \ne i}}^N J_{ij} \sin(\epsilon_j)
\]
We now show that \(\gamma_i\), as defined above, represents the synaptic input to oscillator $i$ from the other connected oscillators in the network. To establish this, we consider the relationship between $\epsilon$ and $\phi$, and the fact that $s=\cos(\phi)$ when \(\phi \in\{0,\pi\}\):
\[
\begin{split}
\gamma_i=-K\sum_{\substack{j=1 \\ j \ne i}}^N J_{ij} \sin(\epsilon_j) &=-K\sum_{\substack{j=1 \\ j \ne i}}^N J_{ij} \sin\left(\phi_j-\frac{\pi}{2}\right)\\
=K\sum_{\substack{j=1 \\ j \ne i}}^N J_{ij} \cos(\phi_j)&=K\sum_{\substack{j=1 \\ j \ne i}}^N J_{ij} s_j
\end{split}
\]

This establishes that \( \gamma_i \) represents the net synaptic input received by oscillator $i$ from the connected oscillators under the conditions described above. Additionally, we note that the self-biasing term can be incorporated by injecting an FHI signal to oscillator with the strength and phase of the signal representing the self-bias input.

The updated state of oscillator $i$ can then be expressed as,
\begin{equation}
\begin{split}
s_i^+ &\approx -\text{sgn}\left[ \tanh(-\gamma_i \Delta t) +\epsilon^{\eta}(\Delta t). \text{sech}^2(\gamma \Delta t) \right]\\\\
&\approx \text{sgn}\left[\tanh\left(K \Delta t\sum_{\substack{j=1 \\ j \ne i}}^N J_{ij} s_j \right) - \vartheta  \right]
\label{s_OIM3}
\end{split}
\end{equation}

\noindent which closely resembles the state update rule for p-bits (Eq.~\eqref{pbit1}), with the factor \( \beta_{\text{eff}} = K \Delta t \) serving as the effective inverse temperature. As detailed in \cite{Wang2021}, the value of K depends on the perturbation projection vector function for the oscillator as well as the amplitude of the perturbation from the oscillators in the network, which can be tuned via the coupling element in the network. This equivalence implies that even in the OIM, the oscillator can approximate Boltzmann sampling thereby enabling the network to function as a p-bit platform.

One of the critical requirements of realizing the above dynamics is to drive the randomly selected oscillator $i$ to $\epsilon_i=0 \:\: \big(\phi_i=\frac{\pi}{2}\big)$. This can be achieved by applying a large FHI signal, \(\text{FHI}^0\), while suppressing the SHI signal $(K_s=0)$. The resulting dynamics can be described by: 
\begin{equation}
\frac{d\epsilon_i}{dt} = -K_{c,i} \sin(\epsilon_i) -K \sum_{\substack{j=1 \\ j \ne i}}^N J_{ij} \sin(\epsilon_i - \epsilon_j) \label{s_OIM4}
\end{equation}

The largest magnitude of the second term is $D_i$---the degree of the node (oscillator) $i$ in the network. By designing $K_{c,i} \gg KD_i$ ensures that the contributions of the second term are small. Consequently, the phase will be driven to $\epsilon_i\approx0$, thereby preparing it for the subsequent stochastic sampling event.\\\\

\noindent \textbf{(ii) Oscillators not being sampled}\\\\
Under steady state, such oscillators have phases $\epsilon =\pm\frac{\pi}{2}$, and the goal is to maintain the configuration when oscillator $i$ is sampled. To achieve this, we apply a strong SHI signal. The corresponding dynamics for an oscillator $j$ that is not being sampled can be then described as follows:

\begin{equation}
\begin{split}
\frac{d\epsilon_j}{dt} = -K \sum_{\substack{k=1 \\ k \ne j}}^N J_{jk} \sin(\epsilon_j - \epsilon_k) + K_s \sin(2\epsilon_j)\\\\
\quad \forall j \in \{1, 2, \dots, i-1, i+1, \dots, N\}
\label{s_OIM1}
\end{split}
\end{equation}

\noindent Furthermore, \(\sin(\epsilon_j-\epsilon_k)\approx0\) since \[
\epsilon_k \in \left\{ -\frac{\pi}{2},\frac{\pi}{2}\right\} \quad \forall j, k \in \{1, 2, \dots, i-1, i+1, \dots, N\}
\]

\noindent Consequently, the dynamics can be reduced to,
\begin{equation}
\begin{split}
\frac{d\epsilon_j}{dt} = -K J_{ji} \sin(\epsilon_j - \epsilon_i) + K_{s} \sin(2\epsilon_j)\\\\
\quad \forall j \in \{1, 2, \dots, i-1, i+1, \dots, N\}
\label{s_OIM1}
\end{split}
\end{equation}

The maximum magnitude of the first term is $|KJ_{ji}|$. Thus, by using $2 K_{s,j} \gg KJ_{ji}$, the phase can be maintained at $\epsilon_j \approx \{-\frac{\pi}{2},\frac{\pi}{2}\}$. These assumptions are further validated through simulations presented in the following section.

Based on the above analysis, algorithm~\ref{alg:algo1} presents the scheme for configuring OIMs as p-bit engines.  Although the analysis above was conducted in a rotated frame of reference, we present the operational scheme in terms of the original phase variable \( \phi \), to maintain consistency with the prevailing conventions in the OIM literature, where the spin states are defined as \( \phi \in \{0, \pi\} \). As previously noted, the relationship between the two frames is given by \( \phi = \frac{\pi}{2} + \epsilon \).

\begin{table}[htbp]

\caption{Operating OIM as a p-bit platform}\label{alg:algo1}
\begin{tabularx}{\linewidth}{@{}r X@{}}
\hline
1: & \textbf{Initialize} all oscillators.\\
2: & \textbf{Apply} SHI signal to all oscillators. The OIM performs gradient
     descent and achieves a steady state characterized by $\phi \in \{0,\pi\}$.\\
3: & \textbf{while} not converged or for a fixed number of iterations \textbf{do}\\
4: & Randomly select an oscillator $i$.\\
   & \{Step (i): Prepare oscillator for stochastic sampling\}\\
5: & Apply $\mathrm{FHI}^{0}$ signal to oscillator $i$ with appropriate amplitude
     and set SHI signal to oscillator $i$ to $0$.\\
   & \{Step (ii): Initiate probabilistic evolution\}\\
6: & Reduce $\mathrm{FHI}^{0}$ signal to $0$.\\
7: & Ramp SHI signal to oscillator $i$.\\
8: & Let $\phi_i$ probabilistically relax to $\phi=0$ or $\phi=\pi$ based on
     synaptic feedback from connected nodes and intrinsic noise.\\
9: & \textbf{end while}\\
\hline\hline
\end{tabularx}
\end{table}

\section{\label{sec:p-bit} Computing with OIM\MakeLowercase{s} configured as P-bit engines}

\subsection{5-Node Adder}

\begin{figure}[h]
    \centering
    \includegraphics[width=1\linewidth]{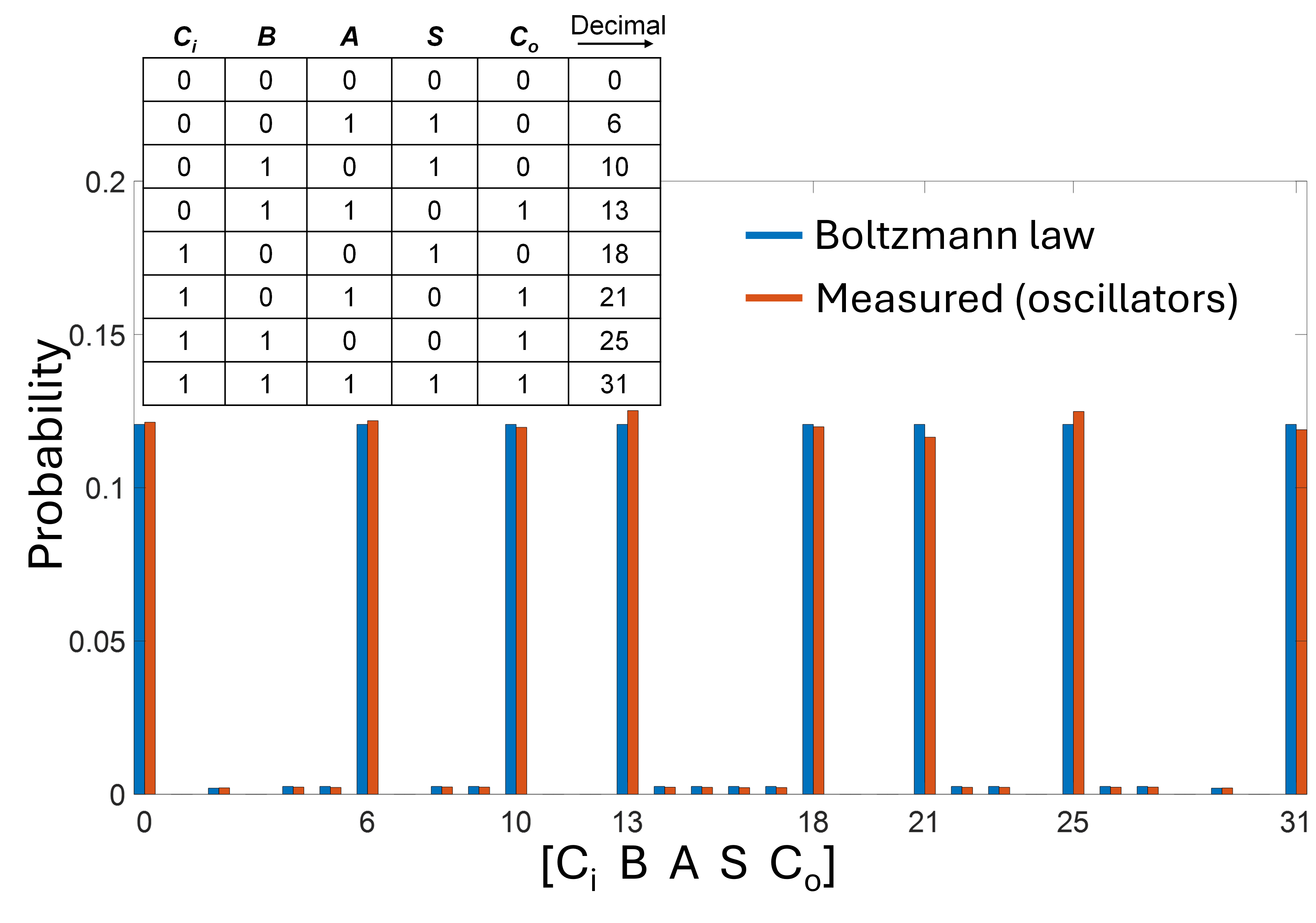}
    \caption{\textbf{Full adder} Probability histogram measured using the oscillators (orange bars, obtained using $5\times10^{5}$ sweeps) compared with the target Boltzmann distribution (blue bars). Following the convention of Ref.~\cite{camsari2017stochastic}, states are indexed by the decimal value of the binary word $[C_{\mathrm{in}}\;A\;B\;S\;C_{\mathrm{out}}]$. The dominant peaks correspond to the valid entries of the full-adder truth table, as highlighted in the inset. The two distributions show excellent agreement, with a measured KL divergence of $7.68 \times 10^{-4}$. Simulation parameters: $K=0.18$; $K_s(t)=K_{s,\text{max}}(1-e^{-\frac{t}{10^{-2}}})$; $K_n=0.1$. }
    \label{fig:sampling}
\end{figure}

To evaluate the sampling capability of our oscillator-based BSN, we benchmark its output distribution against the target Boltzmann distribution for a full adder, following the approach presented by Camsari \textit{et al.}~\cite{camsari2017stochastic}. For this purpose, we construct the corresponding $14 \times 14$ adjacency matrix with the following assignment of nodes: 1–9 represent auxiliary and handle bits, 10 corresponds to $C_{\mathrm{in}}$, 11 to $A$, 12 to $B$, 13 to the sum $S$, and 14 to the carry-out $C_{\mathrm{out}}$. The network is operated in the so-called \emph{truth table mode}, where all input and output terminals are left floating. Under this condition, the distribution obtained using the oscillator-based BSN (using $5 \times 10^{5}$ sweeps) aligns closely with the target Boltzmann distribution, thereby confirming its ability to perform correct stochastic sampling. Here, we note that simulating the analog dynamics of all oscillators for $5 \times 10^{5}$ sweeps (corresponding to a total simulated time of $\sim 2\times10^{7}$ per oscillator) within the stochastic differential equation (SDE) framework (to account for noise) is computationally prohibitive. Therefore, we only simulate the dynamics of oscillator being sampled, while the remaining oscillators are held at their fixed phase values. However, in the following example—computing MaxCut—which requires a smaller number of sweeps, we simulate the entire oscillator network.  

Figure~\ref{fig:sampling} presents the resulting probability histogram, showing an excellent match to the target Boltzmann distribution (KL divergence=$7.68 \times 10^{-4}$). Each state is labeled by the decimal value of the binary string $[\,C_{\mathrm{in}}\;\;A\;\;B\;\;S\;\;C_{\mathrm{out}}\,]$ and the inset highlights the valid truth-table states of the full adder, which appear as the tallest peaks in the distribution.

\subsection{Computing MaxCut}

\begin{figure}
    \centering
    \includegraphics[width=0.9\linewidth]{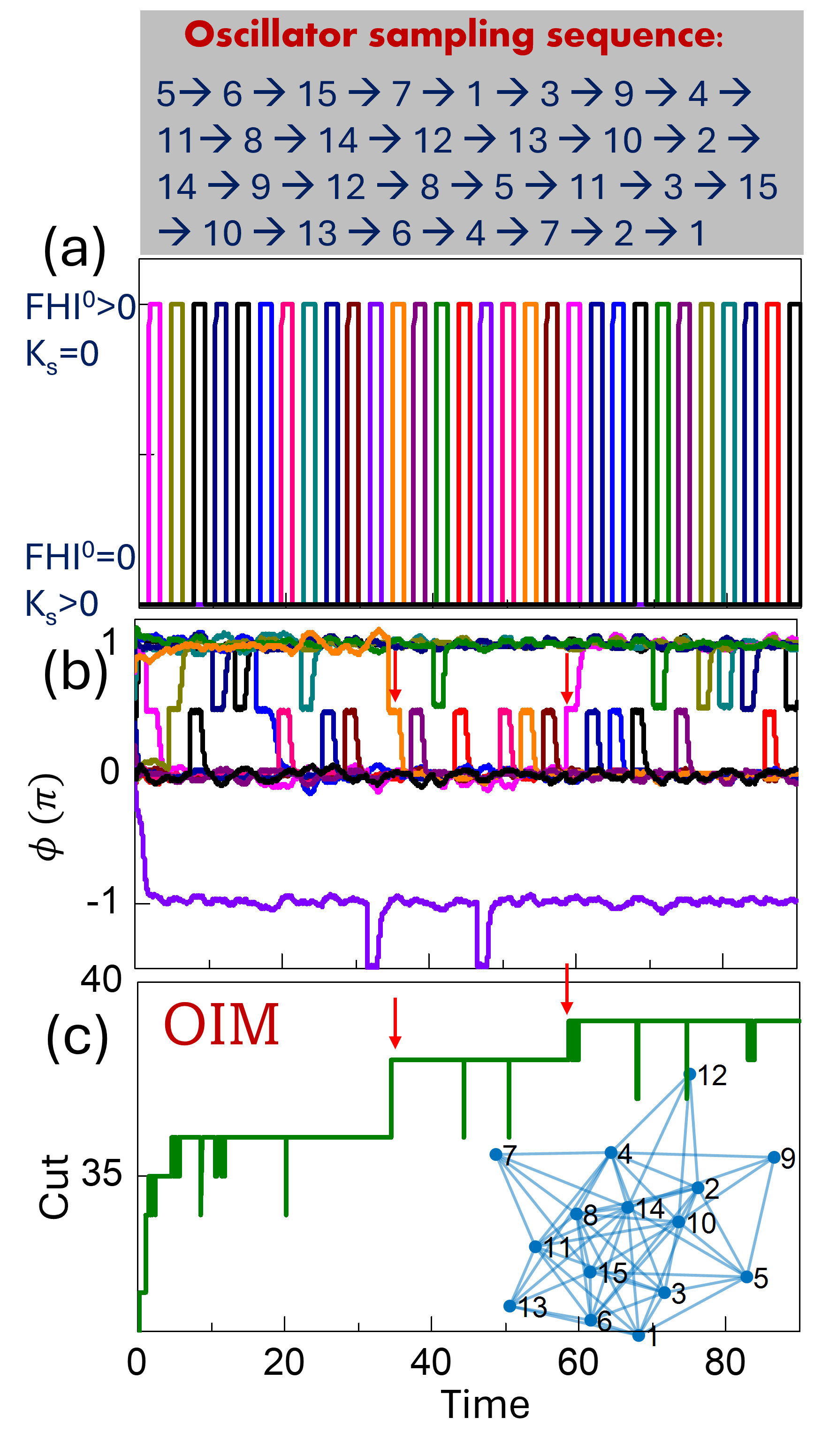}
    \caption{\textbf{Operating OIMs as p-bit platforms.} (a) Randomly generated FHI sequence to the oscillators. The application of FHI to the oscillator is accompanied by the suppression of SHI, and vice-versa. (b) Phase response of the oscillators over time. (c) Evolution of the computed graph cut over time/iterations. With stochastic sampling, the system is able to reach the globally optimal solution (MaxCut =39). The red arrows highlight sampling events that improved cut. \textit{Simulation parameters}: $K=1; K_{s,\text{max}}=2;\: \text{FHI}^0=50;\: K_n \text{ for sampled oscillator}=0.04 $; ramping schedule of the SHI signal: \(K_s(t) = K_{s,\text{max}} \big(1 - e^{-\frac{t}{\tau}}\big)\), where \( K_{s,\text{max}} = 2 \) and \( \tau = 10^{-2} \).}
    \label{fig:Fig.3}
\end{figure}

We now evaluate the archetypal MaxCut problem using the OIM's stochastic sampling mode. Computing the MaxCut of a graph is a NP-hard problem where the objective is to partition the nodes such that the weight of the edges shared among the two sets (i.e., intersect the cut) is maximized. The MaxCut problem directly maps to the solution of the corresponding anti-ferromagnetic Ising Hamiltonian i.e., $J_{ij}=-W_{ij}$, where $\: W_{ij}$ represents the weight of the edges in the graph to be partitioned. While the MaxCut problem has been used as a representative example, the above approach can be applied to other COPs such as graph coloring and finding maximum independent sets, among others.

We demonstrate the functionality using a randomly generated graph with 15 nodes and 59 edges. Figure~\ref{fig:Fig.3}(a) illustrates the sequence of \(\text{FHI}^0\) inputs applied to various randomly sampled oscillators in a sequential manner. The assertion of \(\text{FHI}^0\) is accompanied by suppression of the SHI signal to that oscillator. 
The resulting phase dynamics of the oscillators are shown in Fig.~\ref{fig:Fig.3}(b). It can be observed that during each sampling event,  the sampled oscillator is first driven to a phase of \( \phi_i = \frac{\pi}{2} \) by the application of the \(\text{FHI}^0\) signal, while the phases of the other (non-sampled) oscillators remain at $\phi \in \{0,\pi\}$. Subsequently, the \(\text{FHI}^0\) signal is suppressed, and the  corresponding SHI input is reasserted (not shown in Fig.~\ref{fig:Fig.3}(a) for clarity). As observed in the dynamics, the sampled oscillator then stochastically relaxes toward either \( \phi_i = 0 \) or \( \phi_i = \pi \), with the direction of phase relaxation determined by the competition between the synaptic input and the noise, as noted earlier. Arrows in Fig.~\ref{fig:Fig.3}(b) indicate representative cases where the oscillator flips its state. It is noteworthy that in traditional OIMs, such transitions are likely to have a low probability of occurrence once all oscillator phases have settled to \( \phi_i = 0 \) or \( \phi_i = \pi \) since \( \frac{d\phi}{dt} \), which effectively represents the diving force (\( -\nabla E=  \frac{d\phi}{dt}\)) on the oscillator phase is close to zero for all oscillators i.e., \( \frac{d\phi}{dt} \approx 0 \). Figure~\ref{fig:Fig.3}(c) shows the corresponding evolution of the graph cut. The red arrows in Figs.~\ref{fig:Fig.3}(c) highlight the sampling events that lead to an increase in the graph cut allowing the system to compute the optimal MaxCut. Simulation results on 10 additional randomly generated graphs have been presented in Appendix~\ref{appendix:MaxCut_simulations}.

\begin{figure}[h]
    \centering
    \includegraphics[width=1\linewidth]{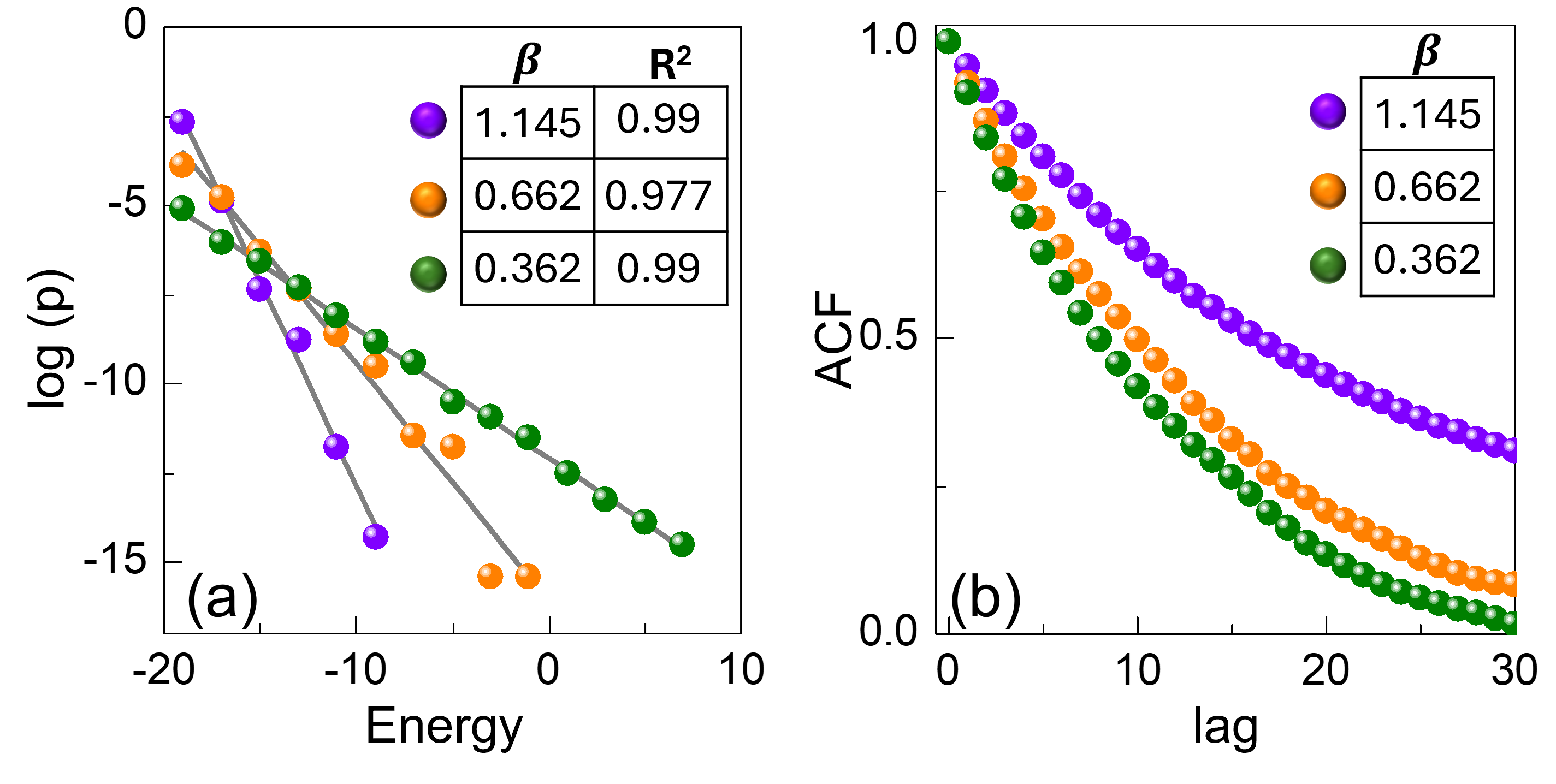}
    \caption{\textbf{Boltzmann sampling behavior.} (a) Plot of $\log $(p) (p: probability) versus energy for varying noise intensities ($K_n = 0.05$, 0.1, 0.15), corresponding to different effective temperatures. The extracted effective inverse temperatures ($\beta$) and the quality of the linear fits ($R^2$) are shown in the inset. (b) Autocorrelation function (ACF) of the system energy as a function of lag, exhibiting a decay toward zero, confirming that the stochastic dynamics effectively decorrelate successive configurations.}
    \label{fig:Fig.MC_BZ}
\end{figure}

While the above simulations demonstrate the application of the OIM's stochastic sampling capability in solving COPs such as MaxCut, we also employ the same example to further corroborate their Boltzmann sampling behavior. Specifically, we perform simulations on the 15-node graph considered in Fig.~\ref{fig:Fig.3} under varying noise intensities, which effectively correspond to different effective temperatures. Similar to the five-state adder analysis, we simulate only the dynamics of the oscillator being sampled, while the remaining oscillator phases are held fixed. Figure~\ref{fig:Fig.MC_BZ}(a) shows the resulting $\log$(p) (p: probability) versus energy distributions obtained over 5000 sweeps. The observed linear dependence confirms that the sampled state probabilities follow the expected Boltzmann relation, with the slope yielding the effective inverse temperature~($\beta$). Furthermore, Fig.~\ref{fig:Fig.MC_BZ}(b) presents the normalized autocorrelation function (ACF)~\cite{BoxJenkins2015} of the system energy as a function of the lag. In all cases, the autocorrelation decays toward zero with increasing delay, indicating that the stochastic dynamics efficiently decorrelate successive configurations.

\begin{figure}[htbp]
    \centering
    \includegraphics[width=0.8\linewidth]{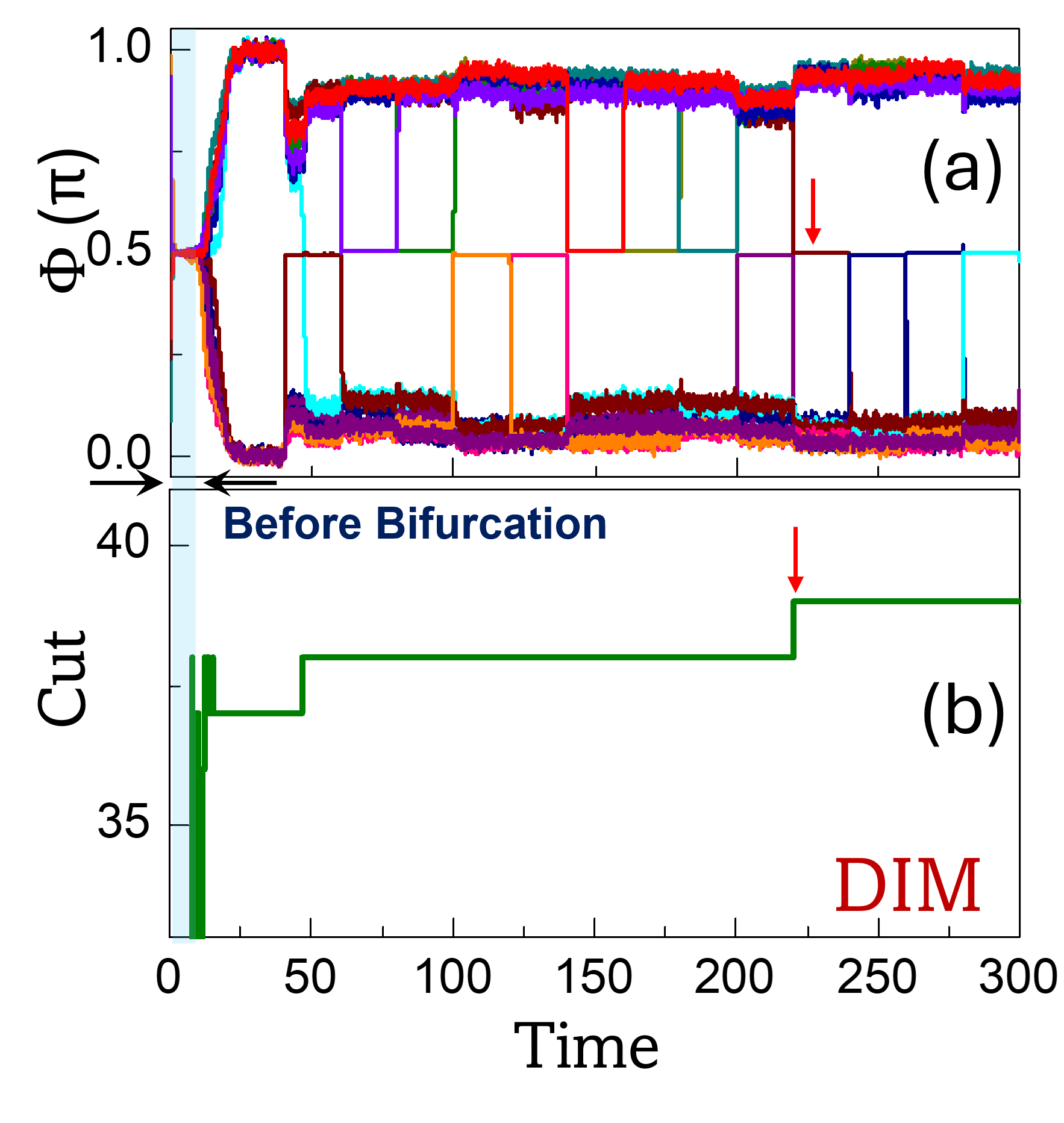}
    \caption{\textbf{Operating DIMs as p-bit platforms.} Evolution of (a) phases in the DIM model; and (b) corresponding graph cut over time. The same graph shown in Fig. \ref{fig:Fig.3} has been considered in the simulation. With stochastic sampling, the system is able to compute the MaxCut (=39) ($K=1; K_{s,\text {max}}=4;\: \: \text{FHI}^0=50;\: \:K_n \text{ for sampled oscillator}=0.006$).}
    \label{fig:Fig.4}
\end{figure}

\section{\label{sec:level1} Realizing P-bit Engines Using Other Analog Ising Machines}

Beyond conventional OIMs, the proposed sampling methodology exhibits potential for generalization to a broader class of analog dynamical systems. As a case in point, we evaluate the implementation of the proposed \textit{sampling mode} in the Dynamical Ising Machine (DIM) recently introduced by the authors ~\cite{cdc9-y234}. Unlike the traditional Kuramoto model, which relies on phase differences, the DIM employs additive phase interactions. The DIM dynamics can be described by:

\begin{equation}
\begin{split}
\frac{d\phi_i}{dt} &= -K \sum_{\substack{j=1 \\ j \ne i}}^N J_{ij} \sin(\phi_i \boldsymbol{+} \phi_j) - K_s \sin(2\phi_i) \label{s_DIM1}
\end{split}
\end{equation}

Specifically, when \( K_s \) is below a certain threshold, the system stabilizes at the trivial state \( \phi = \frac{\pi}{2} \). As \( K_s \) increases beyond this threshold, the system undergoes a bifurcation, leading to the emergence of stable phase configurations at \(\phi \in \{0, \pi\}^N \), which can subsequently be mapped to a spin configuration. The phase dynamics of the DIM, without stochastic sampling, are presented in Appendix \ref{appendix6} for the graph considered in Fig. \ref{fig:Fig.3} . 

While a detailed analysis of the DIM dynamics has been presented in~\cite{cdc9-y234}, it is worth noting that the system exhibits a pitchfork bifurcation, qualitatively similar to that observed in the other popular Ising machine models such as the simulated bifurcation machine (SBM)~\cite{goto2019combinatorial}. This similarity suggests the feasibility of performing stochastic sampling in a broad class of analog dynamical systems beyond the OIM.  

Similar to the OIM, the DIM dynamics in the rotated frame of reference are given by,
\begin{equation}
\begin{split}
\frac{d\epsilon_i}{dt} &=
K \sum_{\substack{j=1 \\ j \ne i}}^N J_{ij} \sin(\epsilon_i + \epsilon_j) + K_s \sin(2\epsilon_i) \\ \\
&= \cos(\epsilon_i)\left(K\sum_{\substack{j=1 \\ j \ne i}}^N J_{ij} \sin( \epsilon_j) \right)
 + K_s \sin(2\epsilon_i)\\\\
&\equiv-\gamma_i\cos(\epsilon_i)\ + K_s \sin(2\epsilon_i) 
 \label{s_DIM2}
\end{split}
\end{equation}
\noindent where $\gamma_i$ has the same definition and meaning as that in the case of the OIM. Moreover, equation~\eqref{s_DIM2} is exactly the same as the corresponding equation~\eqref{s_OIM2} derived for the OIM. Therefore, by employing the same approach and constraints used for the OIM, the updated spin state using the DIM can be derived as follows:

\begin{equation}
\begin{split}
s_i^+ &\approx -\text{sgn}\left[ \tanh(-\gamma_i \Delta t) +\epsilon^{\eta}(\Delta t). \text{sech}^2(\gamma \Delta t) \right]\\\\
&\approx \text{sgn}\left[\tanh\left(K \Delta t\sum_{\substack{j=1 \\ j \ne i}}^N J_{ij} s_j \right) - \vartheta  \right]
\label{s_DIM3}
\end{split}
\end{equation}
Figure~\ref{fig:Fig.4} presents a simulation demonstrating the operation of the DIM in sampling mode to compute the MaxCut of the same graph considered in Fig.~\ref{fig:Fig.3}. As shown in Fig.~\ref{fig:Fig.4}(a), the phase dynamics initially exhibit a bifurcation as \( K_\text{s} \) is ramped up from $K_\text{s}=0$ to $K_\text{s}=4$ (not shown in the figure). At this stage, however, the system has not yet reached the optimal solution. With the onset of stochastic sampling, the system begins to sample the solution space and eventually converges to the optimal MaxCut value of 39, as shown in Fig.~\ref{fig:Fig.4}(b).We note that while both the models exhibit Gibbs sampling, the characteristics (e.g., effective temperature) obtained for a specific set of parameters ($K, K_\text{s}, K_\text{n}$) will depend on the choice of exact dynamics.

\section{\label{sec:level1} Conclusion}
This work builds a conceptual bridge between two paradigms that have traditionally been regarded distinct: analog oscillator-based Ising machines (OIMs) and stochastic sampling-based p-bit engines. By leveraging the natural dynamics of coupled oscillators---specifically through the interplay of SHI and FHI signals---we demonstrate that analog OIMs can perform stochastic sampling without requiring explicit computation of energy functions or the synaptic feedback. 

An intrinsic feature of this approach is the initial phase evolution, during which the oscillator network naturally performs gradient descent that involves the oscillator phases evolving simultaneously. Starting from random initial conditions, the phases converge to discrete states ($\phi_i \approx 0$ or $\phi_i \approx \pi$), effectively settling into a local minimum of the Ising energy landscape. In certain applications such as combinatorial optimization, this simultaneous evolution has the potential to offer a potential speed-up, positioning the system in a low-energy configuration even before the onset of the sampling-mode operation. However, these promising features come with the trade-off of requiring physical connectivity among oscillators---digital p-bit designs are better suited to implement the interaction between p-bits. From an implementation standpoint, this makes the analog approach more appropriate for sparse architectures which aligns well with the operational regimes where traditional p-bit platforms are expected to perform well. 

Nevertheless, it is important to recognize several key factors that must be optimized to ensure reliable operation: \textbf{(a)} The SHI pulse characteristics (amplitude and slew rate) must be carefully designed, as they determine $\Delta t$, which in turn governs the effective temperature, the effective noise, and the validity of the approximations used in Eq.~\eqref{BSN_2} to achieve Gibbs sampling. In practice, a high slew-rate SHI signal (resulting in a small $\Delta t$) is generally desirable to simultaneously satisfy these requirements. \textbf{(b)} Beyond $\Delta t$, the effective temperature also depends on the coupling strength, $K$, among oscillators. Therefore, the coupling strength must be co-designed with $\Delta t$ to realize the desired temperature range, while maintaining frequency locking and satisfying the weak-coupling condition implicit in the dynamics considered here. Since $K$ itself is determined by factors such as the natural frequency, oscillation amplitude, and the magnitude of the physical quantity (e.g., coupling resistance) characterizing the coupling element, these parameters must also be carefully considered to ensure robust operation. \textbf{(c)} Finally, the present analysis does not account for parameter variations or natural frequency mismatches--effects that may play a critical role in determining whether robust operation can be achieved. These design aspects and challenges will need to be systematically investigated going forward. 

From an algorithmic standpoint, we note that since Gibbs sampling is inherently sequential, incorporating techniques such as the graph-coloring method proposed by Niazi \textit{et al.}~\cite{niazi2024training} which enables multiple spins to be updated simultaneously--- need exploration to further enhance performance, and represents a compelling direction for future exploration. Looking ahead, the capability of OIMs—when configured as p-bit engines—to perform stochastic sampling opens promising avenues for neural network applications, including the training of restricted Boltzmann machines.

The overlaps between the analog and probabilistic paradigms identified in this work motivate the extension of this framework to more generalized computational models, including higher-order Ising machines \cite{bashar2023designing,kleyko2023efficient}, p-bits with more than two states \cite{duffee2025extended,whitehead2023cmos}, as well as to other analog systems \cite{goto2019combinatorial,honari2020optical,berloff2025vector,mallick2022computational,delacour2025lagrange,ercsey2011optimization} beyond those explored here.

\section*{ACKNOWLEDGEMENTS}
We gratefully acknowledge Prof. Kerem Camsari for providing valuable insights on the sampling properties of p-bits. This material is based upon work supported in part by ARO award W911NF-24-1-0228 and National Science Foundation grants ($\#$2422333,\,\#2433871).
\vspace{1in}

\renewcommand{\appendixname}{\MakeUppercase{Appendix}}
\appendix

\section{\MakeUppercase{Injection locking in Single Oscillator}}
\addcontentsline{toc}{section}{Injection locking in Single Oscillator} 
\label{appendix1}

We analyze the impact of FHI on oscillator dynamics from an energy-based perspective. For an oscillator driven by an FHI signal at its natural frequency but with a phase offset \( \theta \), the phase dynamics can be described using Adler's equation \cite{adler2006study,bhansali2009gen} as follows:
\begin{equation}
\frac{d\phi}{dt} = -K_c \sin(\phi - \theta)
\label{appendix1-1}
\end{equation}

\noindent The corresponding energy function that the above dynamics minimize can be expressed as:
\begin{equation}
E(\phi) = -K_c \cos(\phi - \theta)
\label{appendix1-2}
\end{equation}

\noindent For this system, the relationship \( -\frac{dE}{d\phi} = \frac{d\phi}{dt} \) holds, implying that:
\[
\frac{dE}{dt} = \frac{dE}{d\phi} \cdot \frac{d\phi}{dt} = -\left( \frac{d\phi}{dt} \right)^2 \leq 0
\]
Thus, the energy monotonically decreases over time, and the system evolves toward a stable fixed point.

The minimum of the energy function occurs at \( \phi^* = \theta \), where \( \frac{dE}{dt} = 0 \). Within the domain \( \phi \in [\theta, \theta + 2\pi) \), the only other fixed point satisfying \( \frac{dE}{dt} = \frac{d\phi}{dt} = 0 \) is at \( \phi^* = \theta + \pi \), which corresponds to a local maximum, and is therefore unstable. For all other values of \( \phi \) in this domain, \( \frac{dE}{dt} < 0 \). Consequently, the oscillator phase converges to the stable fixed point \( \phi = \theta \), although perturbations may be necessary to prevent the dynamics from settling at the unstable fixed point \( \phi^* = \theta + \pi \).

\section{\MakeUppercase{Stability of the Fixed points of the dynamics}}
\label{appendix2}

We analyze the stability of the fixed points associated with the dynamics described in Eq. \eqref{neuron3} of the main text. As previously discussed, the  fixed points of the dynamics are given by, $\epsilon_1^*=\pm\frac{\pi}{2}$ and $\sin(\epsilon_2^*)=\frac{\gamma_i}{2K_s}$.

To investigate the local stability of the system, we analyze the sign of the second derivative:
\[
H(\epsilon_i) = \frac{d^2\epsilon_i}{dt^2} = \gamma_i \sin(\epsilon_i) + 2K_s \cos(2\epsilon_i)
\]

We seek conditions under which \( H(\epsilon_i) < 0 \), indicating local stability.

\begin{itemize}
    \item \textbf{Stability of \( \epsilon_1^* = \frac{\pi}{2} \)}\\\\
    At this point, \( \sin(\epsilon_1^*) = 1 \), \( \cos(2\epsilon_1^*) = -1 \), so:
    \[
    H(\epsilon_1^*) = \gamma_i + (-2K_s) = \gamma_i - 2K_s
    \]
    Hence, \( H(\epsilon_1^*) < 0 \) when \( \gamma_i < 2K_s \).

    \item \textbf{Stability of \( \epsilon_1^* = -\frac{\pi}{2} \)}\\\\
    Here, \( \sin(\epsilon_1^*) = -1 \), \( \cos(2\epsilon_1^*) = -1 \), therefore:
    \[
    H(\epsilon_1^*) = -\gamma_i - 2K_s
    \]
    Thus, \( H(\epsilon_1^*) < 0 \) when \( \gamma_i + 2K_s >0 \).

    \item \textbf{Combined Condition:}\\\\
    Both fixed points \( \epsilon_1^* = \pm \frac{\pi}{2} \) are stable when:
    \[
    \gamma_i^2 < 4K_s^2
    \]

    \item \textbf{Stability of \( \sin(\epsilon_2^*) = \frac{\gamma_i}{2K_s} \)}\\\\
    Substituting into \( H(\epsilon_i) \), we find that:
    \[
    H(\epsilon_2^*) < 0 \quad \text{when} \quad \gamma_i^2 > 4K_s^2
    \]
\end{itemize}

\section{\MakeUppercase{Tuning energy barrier with SHI strength ($K_s$)}}
\label{appendix3}
\begin{figure}[h]
    \centering
    \includegraphics[width=0.8\linewidth]{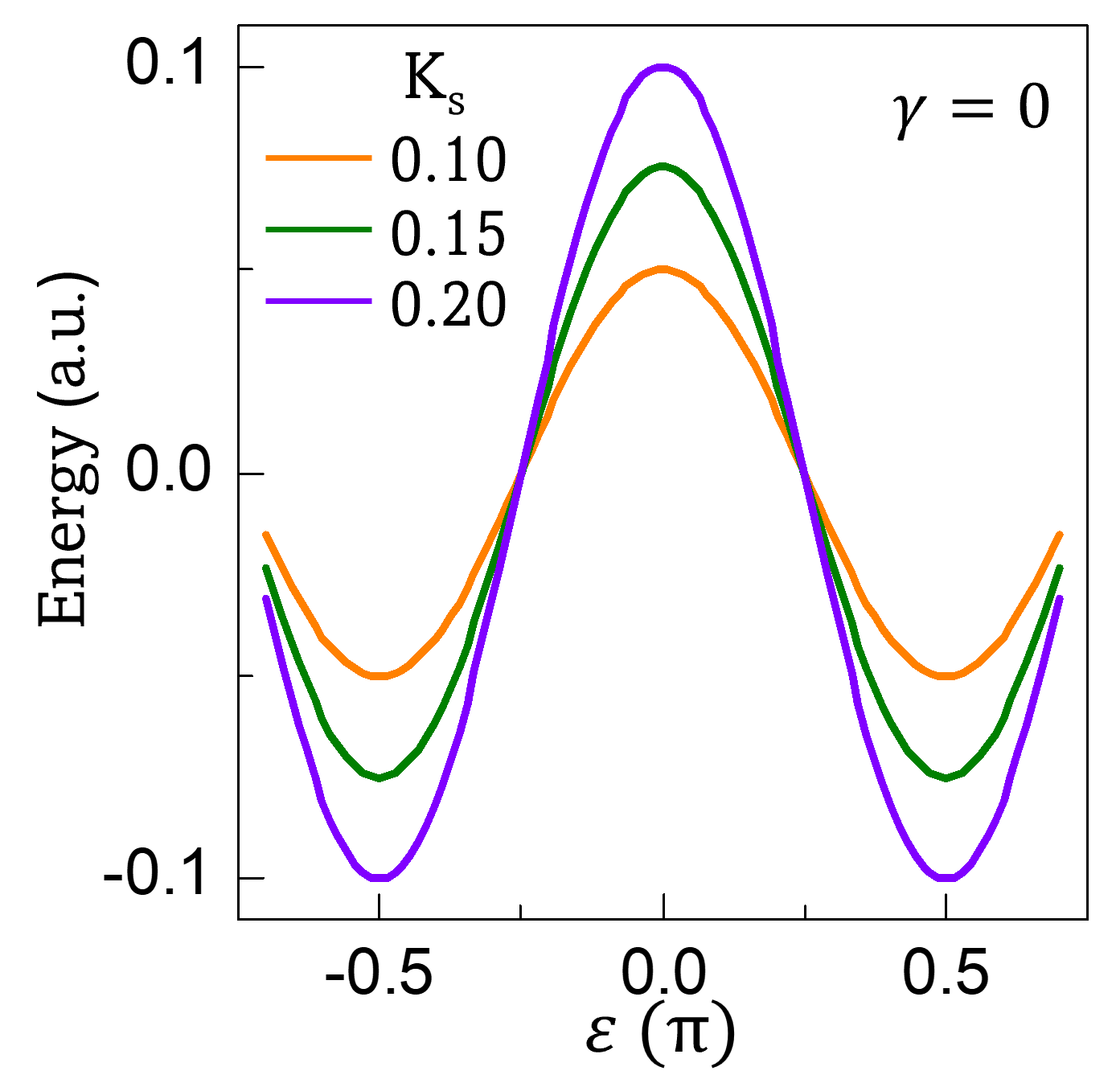}
    \caption{\textbf{Tuning energy barrier with \(\mathbf{K_s}\).} Evolution of the energy barrier with $K_s$}
    \label{fig:Fig.5}
\end{figure}

We analyze the impact of $K_s$ on the energy barrier. The function corresponding to the dynamics described by Eq.~\eqref{neuron3} is,

\begin{equation}
    E=\gamma\sin(\epsilon)+\frac{1}{2}K_s\cos(2\epsilon)
\end{equation}

Figure \ref{fig:Fig.5} shows the resulting energy landscape for different values of $K_s$ ($\gamma=0$). The energy difference between the peak energy (at $\epsilon=0$) and either valley (at $\epsilon=\pm\frac{\pi}{2}$) is given by $(\Delta E)_{max}=|K_s|$.

\section{\MakeUppercase{Temporal Evolution of the Phase}}
\label{appendix4}

We now analyze the dynamics presented in Eq.~\eqref{neuron3}. The implicit analytical solution is given by,
\begin{equation}
    \frac{\zeta(\epsilon) - \gamma \tanh^{-1}(\sin(\epsilon))}{\gamma^2-4K_s^2}=t+C \label{appendix4-1}
\end{equation}

\noindent where,
\begin{equation}
\begin{split}
    \zeta(\epsilon) &= K_s \big( -2 \log(\gamma - 2K_s \sin(\epsilon)) \\
    &+ \log(1 - \sin(\epsilon)) + \log(\sin(\epsilon) + 1) \big)\\                    \notag   
    \end{split}
\end{equation}
and $C$ is the constant of integration. $C=0$ since $\epsilon(t=0)=0$, and $K_s=0$ at $t=0$. 

We now analyze the dynamics under the constraints specified in the main text, which are also restated below for reference: \\\\
\noindent \textbf{(i)} The dynamics are evaluated in the limit $t\rightarrow0$.\\\\
\textbf{(ii)} We model the noise as Gaussian white noise with
\(\langle \eta(t) \rangle = 0, \quad \langle \eta(t)\eta(t') \rangle = 2K_n \delta(t - t'),\) with $K_n$ denoting the noise intensity
i.e., \(\eta(t)\,dt = \sqrt{2K_n}\,dW_t,\) where \(W_t\) is a Wiener process. Equivalently, over a finite timestep $\Delta t$,
\(
\int_{t}^{t+\Delta t} \eta(\tau)\,d\tau \;\sim\; \mathcal{N}\!\big(0,\,2K_n\,\Delta t\big).
\)
\\\\
\textbf{(iii)} At the onset of sampling ($t\rightarrow0$) , we assume \( K_s \rightarrow0\) such that \(K_s\ll |\gamma| \). This reflects the requirement for a low energy barrier in the probabilistic regime. 

\noindent Eq.~\eqref{appendix4-1} can be rearranged to yield,
\begin{equation}
\begin{split}
\sin(\epsilon_i)&=\tanh\left(-\gamma t+\frac{4K_s^2 t+\zeta(\epsilon)}{\gamma}+\epsilon^\eta(t) \right) \\\\ \label{appendix4-2}
&=\tanh\left(-\gamma t+\frac{4K_s^2 t+\zeta(\epsilon)}{\gamma}+\epsilon^\eta(t)\right)
\end{split} \
\end{equation}
Here, $\epsilon^\eta(t)$ represents the perturbation induced by noise. We note that in the above analysis, we approximate the impact of noise as phase jitter; a more exhaustive treatment would model the phase dynamics as a stochastic differential equation.

From the above equation, we group all the terms dependent on $K_s$ as,
\[\begin{split}
&\Theta=\frac{4K_s^2 t+\zeta(\epsilon)}{\gamma}\\\\
\Rightarrow \quad &\sin(\epsilon_i)=\tanh\left(-\gamma t+\Theta+\epsilon^\eta(t)\right)
\end{split}
\]
and evaluate $\Theta$ under the constraints stated above. 

\noindent We begin by simplifying the following terms:\\\\
 $\quad \log(1-\sin(\epsilon))+\log(1+\sin(\epsilon))= \log(\cos^2(\epsilon))$.\\\\
Thus, \( \zeta(\epsilon)=2K_s\left(\log\left(\cos(\epsilon)\right)-\log\left(\gamma-2K_s\sin(\epsilon)\right)\right)\)\\\\

\noindent Next, we apply the following approximations (applicable under the constraints stated above),\\\\
$\bullet$ $\quad \cos(\epsilon)\approx 1-\frac{\epsilon^2}{2} \Rightarrow \log(\cos(\epsilon)) \approx -\frac{\epsilon^2}{2}$\\\\
$\bullet$ $\quad\sin(\epsilon)\approx\epsilon$\\\\
$\bullet$ $\quad \log(\gamma-2K_s\sin(\epsilon))\approx \log(|\gamma|)-\frac{2K_s\epsilon}{\gamma}$\\

\noindent Substituting these approximations into expression for \(\zeta\) yields,
\[
\zeta \approx 2K_s\left(\frac{-\epsilon^2}{2}-\log(|\gamma|)+\frac{2K_s\epsilon}{\gamma} \right)
\]
Subsequently, $\Theta$ can be approximated as,
\[
\begin{split}
\Theta&=\frac{4K_s^2 t+2K_s\left(\frac{-\epsilon^2}{2}-\log(|\gamma|)+\frac{2K_s\epsilon}{\gamma} \right)}{\gamma}\\\\
&=\frac{4K_s^2 t}{\gamma}-\frac{K_s\epsilon^2}{\gamma}-\frac{2K_s\log(|\gamma|)}{\gamma}+\frac{4K_s^2\epsilon}{\gamma^2}
\end{split}
\]

The above expression shows the leading-order behavior of terms dependent on $K_s$ in the phase behavior. Moreover, when $K_s\rightarrow0$ such that $K_s\ll |\gamma|$, \(\Theta\rightarrow0\). Nevertheless it is important that the system parameters be carefully designed to ensure that the dynamics emulate Boltzmann sampling as close as possible. 

\begin{figure*}
   \centering
    \includegraphics[width=0.9\linewidth]{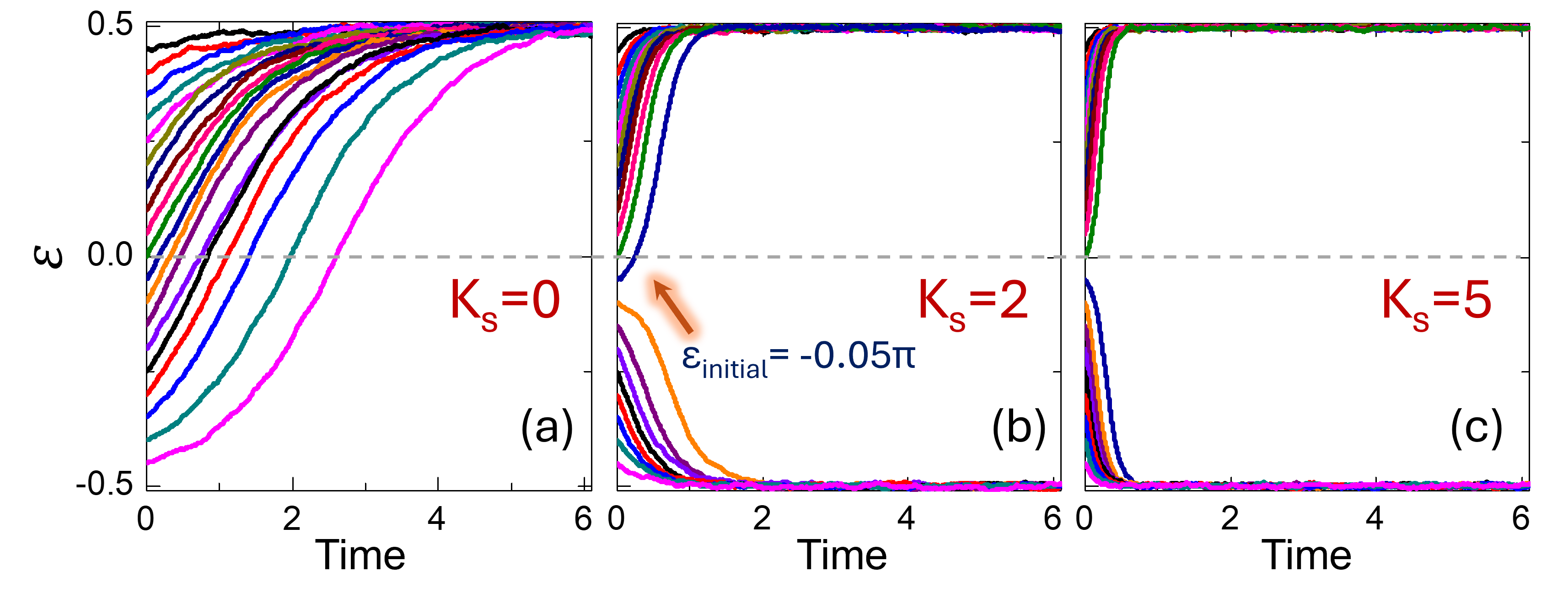}
    \caption{\textbf{Role of SHI in maintaining phase trajectory.} Phase deviation $\epsilon$ as a function of time for: (a) $K_{s}=0$; (b) $K_{s}=2$; (c) $K_{s}=5$. The results in the illustrative example show that a critical SHI  is needed to help the oscillator phase continue to evolve along the direction of its initial perturbation. $|\gamma_1|=1$ has been considered in this example.}
    \label{fig:Fig.6}
\end{figure*}

\noindent Under these conditions, Eq. \eqref{appendix4-2} can be approximated as,
\[\begin{split}
\sin(\epsilon_i)&\approx\tanh\left(-\gamma t+\epsilon^\eta(t) \right)\\\\
&\approx\tanh\left(-\gamma t\right)+\epsilon^\eta(t).\text{sech}^2(\gamma t) \\\\
\end{split}
\]

\noindent Here, $\epsilon^\eta(t)\sim \mathcal{N}\!\big(0,\,2K_n\,t\big)$, and is small such that the $\tanh(.)$ term can be linearized. The noise term, $\epsilon^\eta(t) .\text{sech}^2(\gamma t)$, can be considered as a gaussian distribution representing white noise, and scaled by a function of the synaptic input---sech$^2(\gamma t)$. 
The updated spin state, $s^+=-\text{sgn}\left(\sin(\epsilon^+)\right)$, at a short time instant $\Delta t\rightarrow0$ after sampling has been initiated (at $t=0$), can be expressed as,
\[
\begin{split}
s^+ &\approx \text{sgn}\left[ \tanh(\gamma \Delta t) - \epsilon^\eta(\Delta t) .\text{sech}^2(\gamma \Delta t) \right]\\\\
&\equiv \text{sgn}\left[ \tanh(\gamma \Delta t) - \vartheta \right]
\end{split}
\label{appendix}
\]

\noindent where, $\vartheta\equiv \epsilon^\eta(\Delta t) .\text{sech}^2(\gamma \Delta t)$. Since $\text{sech}^2(\gamma \Delta t) \in (0,1]$, and noise power is assumed to be small, $\vartheta$ has a high probability of being in the interval [-1,+1]. We also note that for $\Delta t \rightarrow0$, $\text{sech}^2(\gamma \Delta t)\rightarrow1 \Rightarrow \vartheta\approxeq \epsilon^\eta(\Delta t)$. Moreover, the gaussian distribution is more representative of the thermal noise found in physical devices.


We also consider the case when both $\gamma$ and $K_s$ are small. Under this assumption, the dynamics for $\epsilon\ll1$ can be approximated as,
\[
\frac{d\epsilon}{dt}\approx -\gamma+2K_s\epsilon  \label{appendix4-3}
\]

Unlike the previous case, using the SDE framework here yields a tractable and elegant solution that offers a clear and intuitive picture of the relative competition between the stochastic and deterministic components.

We begin by considering the linear It\^o SDE
\begin{equation}
\begin{split}
&d\epsilon(t) = \big(-\gamma + 2K_s \epsilon(t)\big)\,dt 
                 + \sqrt{2K_n}\,dW_t, \\\\
&\text{where we note that } \epsilon(0)=0,\; K_s>0 .
\label{eq:sde}
\end{split}
\end{equation}

Integrating factor:
Let $M(t)=e^{-2K_s t}$. Since $M$ is deterministic,
\begin{align}
d\!\big(M(t)\epsilon(t)\big)
 &= M(t)\,d\epsilon(t) + \epsilon(t)\,dM(t) \notag\\
 &= -\gamma M(t)\,dt 
    + \sqrt{2K_n}\,M(t)\,dW_t .
\label{eq:prod}
\end{align}
Integration from $0$ to $t$ yields
\begin{align}
M(t)\epsilon(t) &= -\gamma \int_{0}^{t} M(s)\,ds \notag\\
&\quad + \sqrt{2K_n} \int_{0}^{t} M(s)\,dW_s .
\end{align}
Since 
\[
\int_{0}^{t} e^{-2K_s s}\,ds = \frac{1-e^{-2K_s t}}{2K_s}, 
\quad M(t)^{-1}=e^{2K_s t},
\]
we obtain
\begin{equation}
\epsilon(t) = \frac{\gamma}{2K_s}\,(1 - e^{2K_s t})
 + \sqrt{2K_n}\int_{0}^{t} e^{2K_s (t-s)}\,dW_s .
\label{eq:solution}
\end{equation}

\textbf{Mean:} The stochastic integral in \eqref{eq:solution} has zero mean:
\begin{equation}
\mathbb{E}[\epsilon(t)]
= \frac{\gamma}{2K_s}\,(1 - e^{2K_s t}) .
\label{eq:mean}
\end{equation}

\textbf{Variance:} By It\^o isometry,
\begin{align}
\mathrm{Var}[\epsilon(t)]
 &= 2K_n \int_{0}^{t} e^{4K_s (t-s)}\,ds \notag\\
 &= \frac{K_n}{2K_s}\,\big(e^{4K_s t}-1\big).
\label{eq:var}
\end{align}

The above results imply that for $K_s>0$, both mean and variance diverge exponentially. Thus, no stationary distribution exists. Two limiting cases further clarify the dynamics:

\begin{itemize}
    \item \textbf{Case $\gamma = 0$.}  
    In the absence of the synaptic input, the deterministic contribution reduces to $d\epsilon = 2K_s \epsilon\,dt$.  
    With $\epsilon(0)=0$, this term alone would maintain $\epsilon(t)\equiv 0$.  
    However, in the presence of Gaussian noise, random perturbations are continuously injected and then exponentially amplified by the unstable drift.  
    Consequently, the growth of $\epsilon(t)$ is seeded by stochastic fluctuations and deterministically amplified over time.

    \item \textbf{Case $\gamma \neq 0$.}  
    When $\gamma\neq 0$, both deterministic and stochastic contributions govern the dynamics. As seen from Eq.~\eqref{eq:mean} and Eq.~\eqref{eq:var}, both the drift-induced component and the noise-driven fluctuations diverge as $t$ increases. The relative dominance of these two effects depends on the balance between the deterministic drive, set by $|\gamma|$, and the stochastic forcing, quantified by $K_n$. A stronger deterministic bias leads to more predictable exponential growth, whereas larger noise intensity results in greater dispersion across trajectories.
\end{itemize}

\section{\MakeUppercase{Role of SHI during Stochastic Sampling}}
\label{appendix5}

To analyze how the SHI can help the oscillator maintain the phase trajectory resulting from the stochastic sampling process, we examine Eq.~\eqref{neuron3}:
\[    
\frac{d\epsilon}{dt}=-\gamma\cos(\epsilon)+K_s\sin(2\epsilon) 
\]

As discussed in the main text, the role of SHI is particularly critical when the initial direction of phase relaxation is counter to that expected from synaptic feedback, owing to noise. As noted earlier, the synaptic input (alone) tends to push the phase toward the fixed point opposite to the sign of \(\gamma\). In other words, we seek to analyze the conditions under which the initial flow of the dynamics is towards $\epsilon=+\frac{\pi}{2} (-\frac{\pi}{2})$, even when though the synaptic input is $\gamma>0 \:\:\left(\gamma<0\right)$, respectively.

Assuming that the initial perturbation results in a phase magnitude given by $|\epsilon_{th}|$, the critical condition on $K_s$ to ensure that the phase continues to flow in the same direction can be expressed as,
\begin{equation}
2K_s \sin(|\epsilon_{th}|) > \left|  \gamma \right| 
\label{appendix5-1}    
\end{equation}

This condition ensures the RHS of Eq. ~\eqref{neuron3} maintains the same sign as the phase at $\epsilon=\epsilon_{th}$. Further, since $\sin(.)$ is monotonic in the region, $|\epsilon| \in \{0,\frac{\pi}{2}\}$, the sign of the RHS terms will not change until dynamics reach the corresponding fixed point. We note that in the presence of noise, the above condition should be interpreted in a probabilistic sense.

We also illustrate this behavior using a simple example: a negatively coupled two-oscillator system (with $K=1$) in which the phase of oscillator 2 is fixed at \(\epsilon_2 = -\frac{\pi}{2}\) yielding \(\gamma_1=-1\Rightarrow|\gamma_1|=1\). In this configuration, the energetically favorable state for oscillator 1 is \(\epsilon_1 = \frac{\pi}{2}\), and synaptic feedback is expected to drive the system toward this fixed point. To explore the system's dynamics, oscillator 1 is initialized at various discrete phase values within the range \(\epsilon_{\text{th}} \in [-0.45\pi,\, 0.45\pi]\). Figures~\ref{fig:Fig.6}(a--c) show the evolution of $\epsilon$ for different values of $K_s$. 

In the absence of synaptic hysteresis (i.e., \( K_s = 0 \)), the phase does not preserve the direction of its initial perturbation. Instead, it eventually aligns with the direction of the synaptic input, converging to \( \epsilon = \frac{\pi}{2} \). This behavior is expected, as the inequality in Eq.~\eqref{appendix5-1} is never satisfied in this regime.

Introducing SHI (i.e., $K_s > 0$) enables the oscillator phase to continue evolving along the direction of $\epsilon_{th}$. However, the magnitude of $K_s$ must exceed a certain threshold. This behavior is illustrated in Figs.~\ref{fig:Fig.6}(b) and \ref{fig:Fig.6}(c). In Fig.~\ref{fig:Fig.6}(b), when $K_s$ is below the threshold required for that $\epsilon_{th}=-0.05\pi$, the phase evolution eventually aligns with the direction favored by the synaptic input. In contrast, when $K_s$ is sufficiently, as in Fig.~\ref{fig:Fig.6}(c)), all initial phase values considered here evolve along the direction of their original perturbation. Thus, the SHI injection strength, $K_s$, must be carefully engineered to ensure that one phase magnitudes beyond a certain threshold evolve continue to evolve in the direction of their initial perturbation.

\section{\MakeUppercase{Time Increment $\Delta t$ and Impact of Slew Rate of the SHI signal}}
\label{appendix:slew_rate}
As described in the main text, the infinitesimal time $\Delta t$ is defined as the interval required for the oscillator phase to reach the critical threshold $|\epsilon_{\text{th}}|$, beyond which the phase trajectory cannot reverse and the sampling event is effectively complete. The value of $\Delta t$ is determined by properties of the SHI signal, most notably its slew rate. Qualitatively, this dependence can be understood as follows: a higher slew rate drives the oscillator phase more rapidly toward the fixed points $\epsilon \in \{-\tfrac{\pi}{2}, \tfrac{\pi}{2}\}$, thereby reducing $\Delta t$.  

To provide a quantitative illustration, we consider the SHI signal implemented as a linear ramp with slew rate $\rho$, such that $K_s = \rho t$.  For this case, the phase dynamics (Eq.~\eqref{neuron3}), under the assumption $|\epsilon|\ll 1$, can be expressed as:

\begin{equation}    
\begin{split}
&\frac{d\epsilon}{dt} = -\gamma + (\rho t)(2\epsilon)\\\\
\Rightarrow\: &\frac{d\epsilon}{dt} - 2\rho\, \epsilon\, t= -\gamma \label{eq:slew_rate} 
\end{split}
\end{equation}

where, $\sin(2\epsilon)\sim 2\epsilon$ since $|\epsilon|\ll 1$. 

Using the integrating factor, $\mu(t) = e^{-\rho t^2}$, Eq.~\eqref{eq:slew_rate} can be expressed as,
\begin{equation}    
\frac{d}{dt}\left(\epsilon e^{-\rho t^2}\right) = -\gamma e^{-\rho t^2}
\label{eq:slew_rate_sol} 
\end{equation}

With the initial condition $\epsilon(0)=0$, the resulting phase evolution is described by,

\begin{equation}
\epsilon(t) = -\frac{\sqrt{\pi} \gamma e^{\rho t^2} \, \mathrm{erf}(\sqrt{\rho} t)}{2\sqrt{\rho}} \label{eq:ramp_phase}
\end{equation}

Using Eq.~\ref{eq:ramp_phase}, we compute the time, $\Delta t$ taken by the oscillator phase to reach $|\epsilon_{\text{th}}| \approx \frac{\gamma}{2K_{s,\text{th}}}$ $(|\epsilon_{th} | \ll 1) , \; \text{where} \; K_{s,\text{th}} = \rho (\Delta t)$. Equating Eq.~\eqref{eq:ramp_phase} to $|\epsilon_{th}|$ yields,
\begin{equation}
\begin{split}
    \frac{\gamma}{2\rho (\Delta t)} = \frac{\sqrt{\pi} \gamma e^{\rho (\Delta t)^2} \, \mathrm{erf}(\sqrt{\rho} \Delta t)}{2\sqrt{\rho}}\\\\
    \Rightarrow \sqrt{\rho} \, \Delta t \, \pi \, e^{\rho \Delta t^2} \, \mathrm{erf}(\sqrt{\rho} \Delta t) = 1 \label{eq:SR_Del}
\end{split}    
\end{equation}

Note that the relationship between $\rho$ and $\Delta t$ does not depend on $\gamma$. To express this relationship in terms of $\rho$, we define $y = \sqrt{\rho} \, \Delta t \Rightarrow \rho = \frac{y^2}{(\Delta t)^2}$ where $y$  obeys the relationship,

\begin{equation}
\sqrt{\pi} \, y \, e^{y^2} \, \mathrm{erf}(y) = 1  \label{eq:SR_y}
\end{equation} 
which is Eq.~\eqref{eq:SR_Del} expressed in terms of $y$.

Equation~\eqref{eq:SR_y} implies that $y$ is a constant, whose value can be approximated as $y\approx 0.62$. Thus, the relationship between the ramp rate of the SHI signal and $\Delta t$, can be approximated as,

\begin{equation}
    \begin{split}
    \rho \approx \frac{0.3844}{(\Delta t)^2}\\\\
    \equiv \Delta t \approx \frac{0.62}{\sqrt{\rho}}  \label{eq:SR_Delta_final}
    \end{split}
\end{equation}

Equation~\eqref{eq:SR_Delta_final} shows that $\Delta t$ is inversely proportional to the square root of the slew rate, under the approximations stated above. While this relationship was derived assuming a linear ramp signal for analytical convenience, we expect the same qualitative dependence of $\Delta t$ on $\rho$ to hold for other SHI signal schemes as well.  

\begin{figure}[h]
   \centering
    \includegraphics[width=0.8\linewidth]{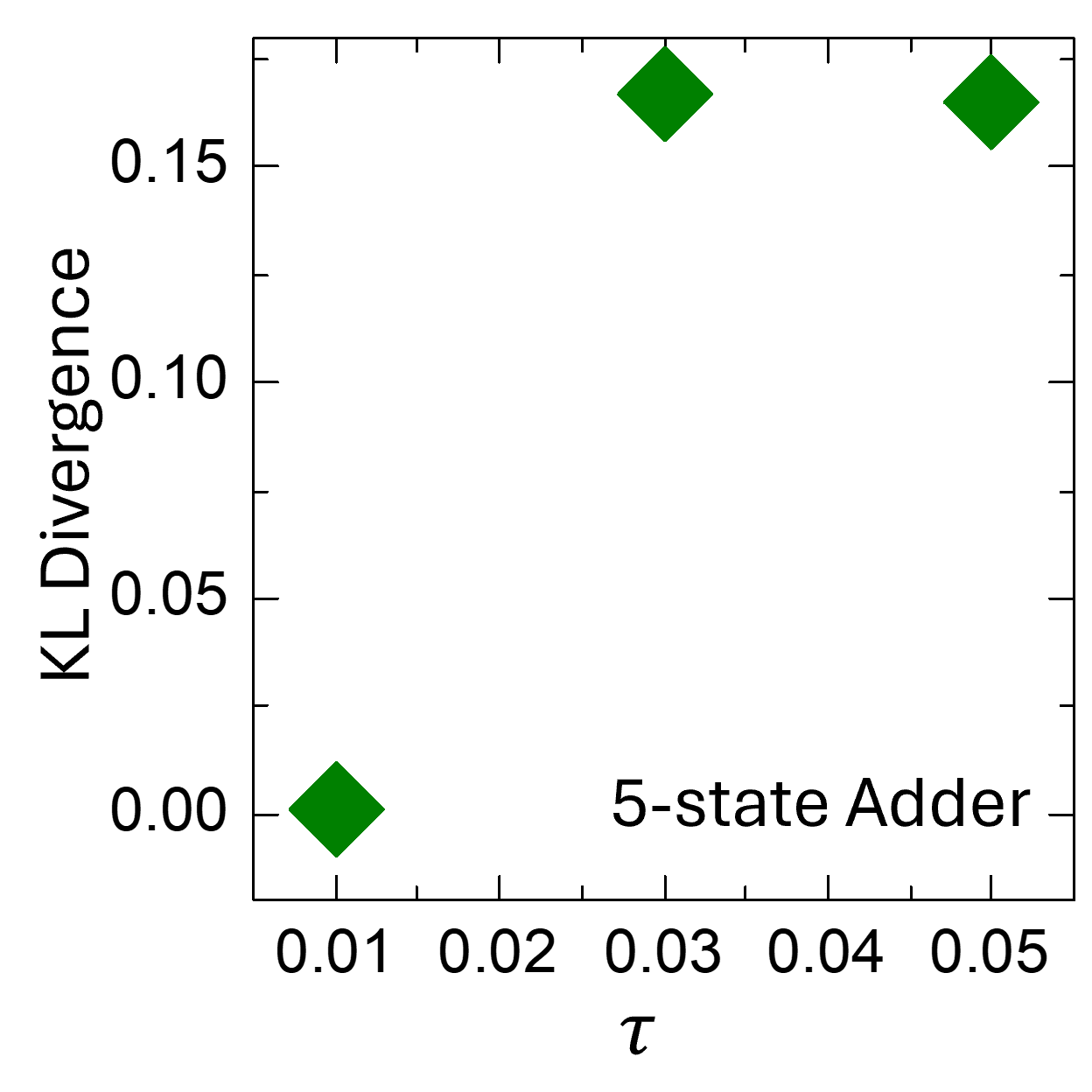}
    \caption{\textbf{Impact of SHI rise time and $\Delta t$.} Measured KL Divergence as a function of $\tau$, where $K_s(t)=K_{s,\text{max}}(1-e^{-\frac{t}{\tau}})$. $\tau$ impacts the time scale over which the SHI signal is asserted.}
    \label{fig:KL}
\end{figure}

The practical implications of the relationship between $\Delta t$ and $\rho$ are:\\\\
\textbf{(a)} $\Delta t$ directly impacts the effective inverse temperature ($\beta$) of the oscillator-based BSN and sets the requirements on the required slew rate.\\\\
\textbf{(b)} Increasing $\Delta t$ suppresses the effect of noise, thereby making the system “less” stochastic.\\
To elucidate this, we consider the expression for the updated spin state given by:
\[
s^+ = \mathrm{sgn}\left[\tanh(\gamma \Delta t) - \epsilon^{\eta}(\Delta t) \, \mathrm{sech}^2(\gamma \Delta t)\right]
\]

The noise term, \(\epsilon^{\eta}(\Delta t) \, \mathrm{sech}^2(\gamma \Delta t)\), indicates that the noise perturbation is scaled by $\text{sech}^2(\gamma \Delta t)$ which has a maximum value (=1) when $\gamma \Delta t =0$, and diminishes (with the value approaching 0) as the value of the argument of the $\text{sech}^2(.)$ increases. Consequently, a large value of $\Delta t$ (for a given $\gamma$), effectively diminishes the impact of noise making the system less stochastic.

We evaluate the impact of the time-scale over which the SHI signal is asserted (and consequently, $\Delta t$) on the sampling properties. Fig.~\ref{fig:KL} shows the measured KL divergence parameter for a 5-state adder (considered above) as a function of $\tau$-- the time-constant associated with the SHI signal; $K_s(t)=K_{s,\text{max}}(1-e^{-\frac{t}{\tau}})$. As noted above, this also impacts $\Delta t$. Consistent with the preceding analysis, an increase in $\tau$ leads to a corresponding rise in the KL divergence, indicating degraded sampling quality at longer $\tau$  durations.

\section{\MakeUppercase{Phase configuration of Non-Sampled Oscillators}}
\label{appendix:non_sampled_oscillators}

As noted in the main text, it is important to ensure that the phases of the non-sampled oscillators are maintained at\[
\epsilon_j \in \left\{ -\tfrac{\pi}{2}, \tfrac{\pi}{2} \right\} 
\;\;\equiv\;\; 
\phi_j \in \{0, \pi\}, 
\quad \forall\, j \in \{1,2,\dots,N\} \setminus \{i\},
\] 
\noindent (where, $i$ is the sampled oscillator) to ensure that the oscillator dynamics map to Gibbs sampling. To explain this requirement, we being with Eq.~\eqref{s_OIM1a}, which is rewritten below for reference:

\begin{equation}
\frac{d\epsilon_i}{dt} = -K \sum_{\substack{j=1 \\ j \ne i}}^N J_{ij} \sin(\epsilon_i - \epsilon_j) + K_s \sin(2\epsilon_i)
\label{eq:ap_sOIM1}
\end{equation}

Expanding the $\sin(\epsilon_i-\epsilon_j)$ term, Eq.~\eqref{eq:ap_sOIM1} can be expressed as,
\begin{equation}
\begin{split}
&\frac{d\epsilon_i}{dt} = \\
&-K \bigg(-\cos(\epsilon_i)\sum_{\substack{j=1 \\ j \ne i}}^N J_{ij} \sin(\epsilon_j)
+ \sin(\epsilon_i)\sum_{\substack{j=1 \\ j \ne i}}^N J_{ij} \cos(\epsilon_j)\bigg)\\
&+ K_s \sin(2\epsilon_i)
\label{eq:ap_sOIM2}
\end{split}
\end{equation}

When $\epsilon_j \in \{-\frac{\pi}{2},\frac{\pi}{2}\} \equiv \phi_j \in\{0,\pi\}$,
\[
\sin(\epsilon_i)\sum_{\substack{j=1 \\ j \ne i}}^N J_{ij} \cos(\epsilon_j)=0
\] reducing Eq.~\eqref{eq:ap_sOIM2} to Eq.~\eqref{s_OIM2} in the main text, which in turn establishes that the OIM dynamics can perform Gibbs sampling.

In contrast, when $\epsilon_j \notin \{-\frac{\pi}{2},\frac{\pi}{2}\} \equiv \phi_j \notin\{0,\pi\}$, then 
\[
\sin(\epsilon_i)\sum_{\substack{j=1 \\ j \ne i}}^N J_{ij} \cos(\epsilon_j)\neq 0
\] 
which introduces an additional component \emph{orthogonal} to $\cos(\epsilon_i)$ that breaks the direct Gibbs mapping (except in special cases where $\sum_{\substack{j=1 \\ j \ne i}}^N J_{ij} \cos(\epsilon_j)$ vanishes or averages to zero).

\section{\MakeUppercase{Additional MaxCut results}}
\label{appendix:MaxCut_simulations}

\begin{figure}[h]
   \centering
    \includegraphics[width=1\linewidth]{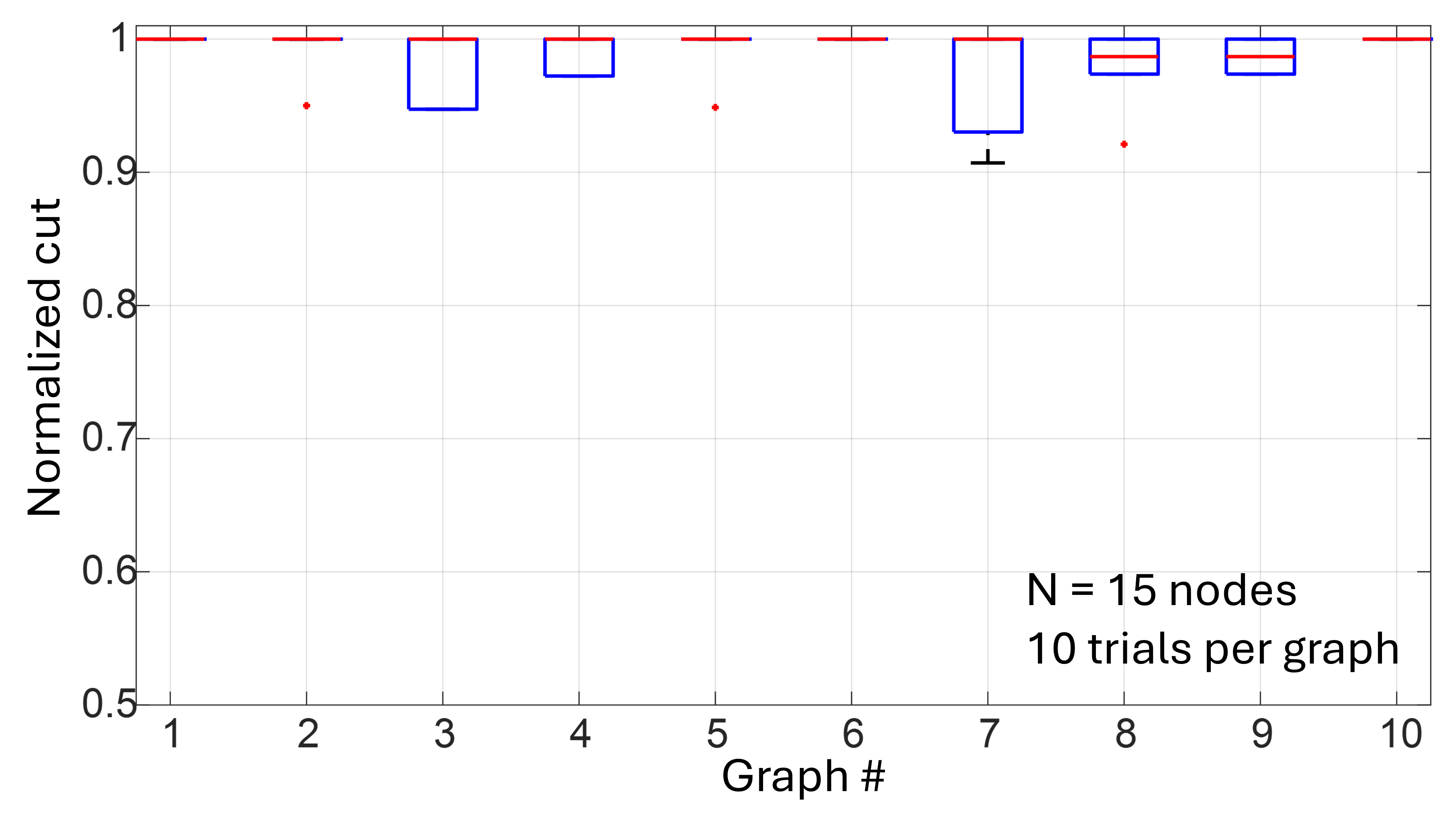}
    \caption{\textbf{Computing MaxCut using OIM-based p-bit engine.} Box plot showing distribution of graph cuts obtained for 10 randomly generated 15-node graphs. Each graph is simulated 10 times. The SHI scheme is the same as that used in the main text.}
    \label{fig:MC}
\end{figure}

We present the MaxCut results for ten additional randomly generated 15-node graphs. Figure~\ref{fig:MC} shows the distribution of the obtained cut values using a box plot, where each cut value is normalized with respect to the corresponding Maximum Cut. Each graph is simulated 10 times.

\section{\MakeUppercase{Dynamics of Dynamical Ising Machine}}
\label{appendix6}

\begin{figure}[h]
   \centering
    \includegraphics[width=1\linewidth]{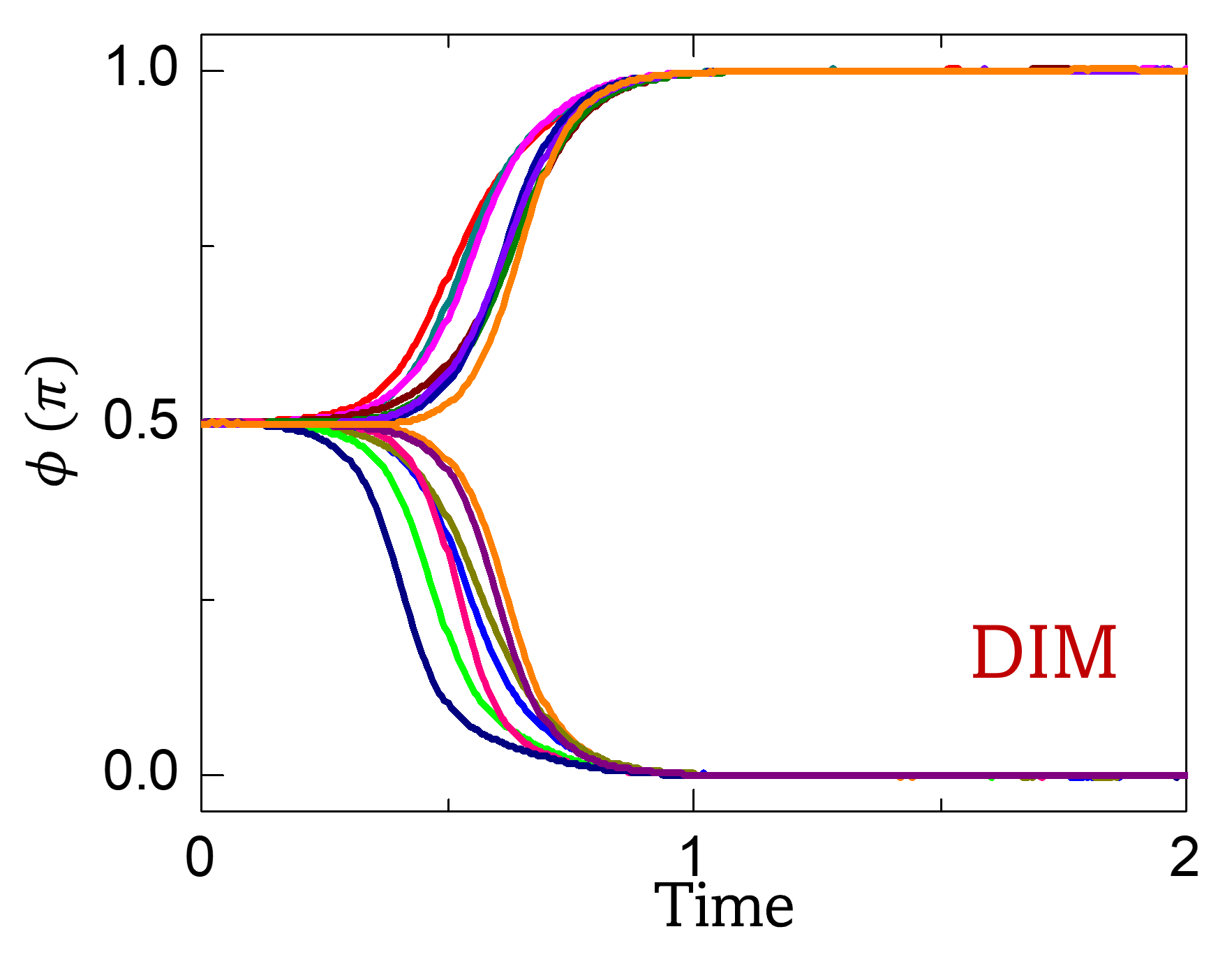}
    \caption{\textbf{Dynamical Ising Machine.} Evolution of $\phi$ in the DIM model for the graph considered in Fig.~\ref{fig:Fig.3} (K=1; $K_s(t)=0.4t$).}
    \label{fig:Fig.7}
\end{figure}

Figure \ref{fig:Fig.7} evaluates the graph considered in Fig. \ref{fig:Fig.3} using the analog dynamics of the DIM (without stochastic sampling). A bifurcation similar to that exhibited by other models such as SBM \cite{goto2019combinatorial} is observed.

\def\bibsection{\section*{References}}  
\bibliography{main_10} 
\end{document}